\documentclass[showpacs,amsmath,amssymb,aps,prd,superscriptaddress,nofootinbib,twocolumn]{revtex4}

\usepackage{graphicx}
\usepackage{amsmath}
\usepackage{multirow}

\usepackage{latexsym}
\usepackage{envmath}

\def\prl{Phys. Rev. Lett.}
\def\prd{Phys. Rev. D}
\def\cqg{Class. Quantum Grav.}

\def\pr{Phys. Rev.}

\def\Gammaflat{\hat \Gamma}
\def\metricflat{\hat \gamma}
\def\Dflat{\hat {\mathcal D}}
\def\part_n{\partial_\perp}

\begin{document}
   
\title{Numerical Relativity in Spherical Polar Coordinates:\\
Evolution Calculations with the BSSN Formulation}

\author{Thomas W. Baumgarte}
\affiliation{Max-Planck-Institute f{\"u}r Astrophysik,
Karl-Schwarzschild-Str.~1, D-85748, Garching bei M{\"u}nchen, Germany}
\affiliation{Bowdoin College, Brunswick, ME 04011, USA}

\author{Pedro J. Montero}
\affiliation{Max-Planck-Institute f{\"u}r Astrophysik,
Karl-Schwarzschild-Str.~1, D-85748, Garching bei M{\"u}nchen, Germany}

\author{Isabel Cordero-Carri\'{o}n}
\affiliation{Max-Planck-Institute f{\"u}r Astrophysik,
Karl-Schwarzschild-Str.~1, D-85748, Garching bei M{\"u}nchen, Germany}

\author{Ewald M\"{u}ller}
\affiliation{Max-Planck-Institute f{\"u}r Astrophysik,
Karl-Schwarzschild-Str.~1, D-85748, Garching bei M{\"u}nchen, Germany}

\begin{abstract}
In the absence of symmetry assumptions most numerical relativity simulations adopt Cartesian coordinates.   While Cartesian coordinates have some desirable properties, spherical polar coordinates appear better suited for certain applications, including gravitational collapse and supernova simulations.   Development of numerical relativity codes in spherical polar coordinates has been hampered by the need to handle the coordinate singularities at the origin and on the axis, for example by careful regularization of the appropriate variables.  Assuming spherical symmetry and adopting a covariant version of the BSSN equations, Montero and Cordero-Carri\'on recently demonstrated that such a regularization is not necessary when a partially implicit Runge-Kutta (PIRK) method is used for the time evolution of the gravitational fields.  Here we report on an implementation of the BSSN equations in spherical polar coordinates without any symmetry assumptions.  Using a PIRK method we obtain stable simulations in three spatial dimensions without the need to regularize the origin or the axis.  We perform and discuss a number of tests to assess the stability, accuracy and convergence of the code, namely weak gravitational waves, ``hydro-without-hydro" evolutions of spherical and rotating relativistic stars in equilibrium, and single black holes.
\end{abstract}


\maketitle

\section{Introduction}
The first announcements of successful binary black hole simulations \cite{Pre05b,CamLMZ06,BakCCKM06a} marked an important break-through in numerical relativity and triggered a burst of activity in the field.  While most current simulations adopt some variation of the BSSN formulation \cite{NakOK87,ShiN95,BauS98} together with what have become ``standard coordinates" (namely 1+log slicing \cite{BonMSS95} and the ``Gamma-driver" condition \cite{AlcBDKPST03}), different implementations differ in many details.  Most current, three-dimensional numerical relativity codes share one feature, though, namely Cartesian coordinates.  While Cartesian coordinates have many desirable properties, there are applications, for example gravitational collapse and supernova calculations, for which spherical polar coordinates would be better suited.\footnote{Cartesian or spherical polar coordinates are not the only two possibilities, of course.  In particular, multi-patch applications, combining some of the advantages of both, may present an attractive alternative at least for some applications (see, e.g., \cite{SchDDT06} for an implementation in numerical relativity).}

Implementing a numerical relativity code in spherical polar coordinates poses several challenges.  The first challenge lies in the equations themselves.  The original version of the BSSN formulation, for example, explicitly assumes Cartesian coordinates (by assuming that the determinant of the conformally related metric be one).   This issue has been resolved by Brown \cite{Bro09}, who introduced a covariant formulation of the BSSN equations that is well-suited for curvilinear coordinate systems (compare \cite{BonGGN04}).

Another challenge is introduced by the coordinate singularities at the origin and the axis, which introduce singular terms into the equations.  Although the regularity of the metric ensures that, analytically, these terms cancel exactly, this is not necessarily the case in numerical applications, and special care has to be taken in order to avoid numerical instabilities.

Several methods have been proposed to enforce regularity in curvilinear coordinates.  One possible approach is to rely on a specific gauge, e.g.~polar-areal gauge~\cite{BarP83,Cho91}, together with a suitable choice of the dynamical variables.  Numerous different such methods have been implemented in spherical symmetry (see, e.g., \cite{BauS10} for an overview); examples in axisymmetry include \cite{BarP83,Eva86,ShaT92}.  This approach has some clear limitations.  It is not obvious how to generalize these methods to relax the assumption of axisymmetry; moreover the restriction of the gauge freedom prevents adoption of the ``standard gauge" that proved to be successful in evolutions with the BSSN formulation.

An alternative method is to apply a regularization procedure, by which both the appropriate parity regularity conditions and local flatness are enforced in order to achieve the desired regularity of the evolution equations (see \cite{ChoHLP03,Rin05,Alc05,RuiTAN08,Rin09,Alc11,Sor10} for examples).  Typically, these schemes involve the introduction of auxiliary variables as well as finding evolution equations for these variables.  The resulting schemes are quite cumbersome, which may explain why, to the best of our knowledge, no such scheme has been implemented without any symmetry assumptions.

In yet an alternative approach, Cordero-Carri\'on {\it et.al.} \cite{CorCDI12} recently adopted a partially implicit Runge-Kutta (PIRK) method  (see also \cite{AscRW95}) to evolve the hyperbolic, wave-like equations in the Fully Constrained formulation of  the Einstein equations (see \cite{BonGGN04}).  Essentially, PIRK methods evolve regular terms in the evolution equations explicitly, and then use these updated values to evolve singular terms implicitly (see \cite{CorCD12} and Section \ref{sec:PIRK} below for details).  Following this success, Montero \& Cordero-Carri\'on \cite{MonC12}, assuming spherical symmetry, applied a second-order PIRK method to the full set of the BSSN Einstein equations in curvilinear coordinates, and produced the first successful numerical simulations of vacuum and  non-vacuum spacetimes using the covariant BSSN formulation in spherical coordinates without the need for a regularization algorithm at the origin (or without performing a spherical reduction of the equations, compare \cite{Gar08,Bern10}).

In this paper we present a new numerical code that solves the BSSN equations in three-dimensional spherical polar coordinates without any symmetry assumptions.  The code uses a second-order PIRK method to integrate the evolution equations in time. This approach has the additional advantage that it imposes no restriction on the gauge choice.   We consider a number of test cases to demonstrate that it is possible to obtain stable and robust evolutions of axisymmetric and non-axisymmetric spacetimes without any special treatment at the origin or the axis.

The paper is organized as follows.  In Section \ref{basic_equations} we present the basic equations; we will review the covariant formulation of the BSSN equations, and will then specialize to spherical polar coordinates.  In Section \ref{sec:numerics} we will briefly review PIRK methods and will then describe other specifics of our numerical implementation.  In Section \ref{sec:numerical_examples} we present numerical examples, namely weak gravitational waves, ``hydro-without-hydro'' simulations of static and rotating relativistic stars, and single black holes.  Finally we summarize and discuss our findings in Section \ref{sec:discussion}.  We also include two appendices; in Appendix \ref{appendixA} we describe an analytical form of the flat metric in spherical polar coordinates that provides a useful test of the numerical implementation of curvature quantities, while in Appendix \ref{appendixB} we list the specific source terms for our PIRK method applied to the BSSN equations.

Throughout this paper we use geometrized units in which $c=G=1$.  Indices $a, b, \ldots$ denote spacetime indices, while $i, j, \ldots$ represent spatial indices.

\section{Basic equations}
\label{basic_equations}

\subsection{The BSSN equations in covariant form}

We adopt Brown's covariant form \cite{Bro09} of the BSSN formulation \cite{NakOK87,ShiN95,BauS98}.  In particular, we write the conformally related spatial metric
$\bar \gamma_{ij}$ as
\begin{equation}
\bar \gamma_{ij} = e^{-4 \phi} \gamma_{ij},
\end{equation}
where $\gamma_{ij}$ is the physical spatial metric, and $e^\phi$ a conformal factor.  In the original BSSN formulation the determinant $\bar \gamma$ of the conformally related metric is fixed to unity, which completely determines the conformal factor.  This approach is suitable when Cartesian coordinates are used, but not in more general coordinate systems.  We will pose a different condition on $\bar \gamma$ below, but note already that
\begin{equation}
e^{4 \phi} = \left( \bar \gamma / \gamma \right)^{1/3}.
\end{equation}
The advantage of this approach is that all quantities in this formalism may be treated as tensors of weight zero (see also \cite{BonGGN04}).  We also denote 
\begin{equation}
\bar A_{ij} = e^{-4 \phi} \left( K_{ij} - \frac{1}{3} \gamma_{ij} K \right)
\end{equation}
as the conformally rescaled extrinsic curvature.  Slightly departing from Brown's approach we assume this quantity to be trace-free, while Brown allows $\bar A_{ij}$ to have a non-zero trace.  In the above expression $K_{ij}$ is the physical extrinsic curvature and $K = \gamma^{ij} K_{ij}$ its trace.

Introducing a background connection $\Gammaflat^i_{jk}$ (compare \cite{BonGGN04}) we now define
\begin{equation} \label{deltagamma}
\Delta \Gamma^{i}_{jk} = \bar \Gamma^i_{jk} - \Gammaflat^i_{jk}
\end{equation}
which, unlike the two connections themselves, transform as a tensor field.  We also define the trace of these variables as
\begin{equation}
\Delta \Gamma^i \equiv \bar \gamma^{jk} \Delta \Gamma^i_{jk}.
\end{equation}
It is not necessary for the background connection to be associated with any metric.  In Section \ref{sec:implementation} below we will specialize to applications in spherical polar coordinates and hence will assume that the $\Gammaflat^i_{jk}$ are associated with the flat metric in spherical polar coordinates.  This assumption affects the equations in the remainder of this Section in only one way, namely, we will assume that the Riemann tensor associated with the connection $\Gammaflat^i_{jk}$ vanishes, as is appropriate when the background metric is flat.

Finally, we define the connection vector $\bar \Lambda^i$ as a new set of independent variables that are equal to the $\Delta \Gamma^i$ when the constraint
\begin{equation}
{\mathcal C}^i \equiv \bar \Lambda^i - \Delta \Gamma^i = 0
\end{equation}
holds.  The vector $\bar \Lambda^i$ plays the role of the ``conformal connection functions" $\bar \Gamma^i$ in the original BSSN formulation, but, unlike the $\bar \Gamma^i$, the $\bar \Lambda^i$ transform as a rank-1 tensor of weight zero (compare exercise 11.3 in \cite{BauS10}).  In the following we will evolve the variables $\bar \Lambda^i$ as independent variables, satisfying their own evolution equation.  

In order to determine the conformal factor $e^{\phi}$ we specify the time evolution of the determinant of the conformal metric.  In this paper we adopt Brown's ``Lagrangian" choice 
\begin{equation} \label{dgammadt}
\partial_t \bar \gamma = 0.
\end{equation}  
Defining
\begin{equation}
\part_n \equiv \partial_t - {\mathcal L}_\beta,
\end{equation}
where ${\mathcal L}_\beta$ denotes the Lie derivative along the shift vector $\beta^i$, we then obtain the following set of evolution equations 
\begin{subequations} \label{evolution}
\begin{eqnarray}
\part_n \bar \gamma_{ij} & = & - \frac{2}{3} \bar \gamma_{ij} \bar D_k \beta^k - 2 \alpha \bar A_{ij} \\
\part_n \bar A_{ij} & = & - \frac{2}{3} \bar A_{ij} \bar D_k \beta^k - 2 \alpha \bar A_{ik} \bar A^k {}_j 
+ \alpha \bar A_{ij} K \nonumber \\
&& + e^{- 4 \phi} \Big[ - 2 \alpha \bar D_i \bar D_j \phi + 4 \alpha \bar D_i \phi \bar D_j \phi \nonumber \\
& & ~~~~~~~ + 4 \bar D_{(i} \alpha \bar D_{j)} \phi - \bar D_i \bar D_j \alpha 
\nonumber\\ 
& & ~~~~~~~ + \alpha (\bar R_{ij} - 8\pi S_{ij}) \Big]^{\rm TF} \\
\part_n \phi & = & \frac{1}{6} \bar D_k \beta^i - \frac{1}{6} \alpha K \\
\part_n K & = & \frac{\alpha}{3} K^2 + \alpha \bar A_{ij} \bar A^{ij} 
- e^{- 4 \phi}  ( \bar D^2 \alpha + 2 \bar D^i \alpha \bar D_i \phi ) \nonumber \\
& & + 4 \pi \alpha (\rho + S) \\
\part_n \bar \Lambda^i & = & \bar \gamma^{jk} \Dflat_j \Dflat_k \beta^i 
+ \frac{2}{3} \Delta \Gamma^i \bar D_j \beta^j + \frac{1}{3} \bar D^i \bar D_j \beta^j \nonumber \\
&  & - 2 \bar A^{jk} ( \delta^i{}_j \partial_k \alpha - 6 \alpha \delta^i{}_j \partial_k \phi
- \alpha \Delta \Gamma^i_{jk} ) \nonumber \\
& & - \frac{4}{3} \alpha \bar \gamma^{ij} \partial_j K - 16 \pi \alpha \bar \gamma^{ij} S_j.
\end{eqnarray}
\end{subequations}
(compare equations (21) in \cite{Bro09}).  In the above equations, $\alpha$ is the lapse function, $\Dflat_i$ denotes a covariant derivative that is built from the background connection $\Gammaflat^i_{jk}$ (and hence, in our implementation, associated with the flat metric $\metricflat_{ij}$ in spherical polar coordinates) and the superscript ${\rm TF}$ denotes the trace-free part.  The matter sources $\rho$, $S_i$, $S_{ij}$ and $S$ denote the density, momentum density, stress, and the trace of the stress as observed by a normal observer, respectively, and are defined by
\begin{subequations}
\begin{eqnarray}
\rho & \equiv & n_a n_b T^{ab}, \\
S_i & \equiv & - \gamma_{ia} n_b T^{ab}, \\
S_{ij} & \equiv & \gamma_{ia} \gamma_{jb} T^{ab}, \\
S & \equiv & \gamma^{ij} S_{ij}.
\end{eqnarray}
\end{subequations}
Here
\begin{equation}
n_a = (-\alpha,0,0,0)
\end{equation}
is the normal one-form on a spatial slice, and $T^{ab}$ is the stress-energy tensor.

We compute the Ricci tensor $\bar R_{ij}$ associated with $\bar \gamma_{ij}$ from
\begin{eqnarray} \label{ricci}
\bar R_{ij} & = & - \frac{1}{2} \bar \gamma^{kl} \Dflat_k \Dflat_l \bar \gamma_{ij} + 
\bar \gamma_{k(i} \Dflat_{j)} \bar \Lambda^k + \Delta \Gamma^k \Delta \Gamma_{(ij)k} \nonumber \\
& & + \bar \gamma^{kl} \left( 2 \Delta \Gamma^m_{k(i} \Delta \Gamma_{j)ml} 
+ \Delta \Gamma^m_{ik} \Delta \Gamma_{mjl} \right).
\end{eqnarray}
In all of the above expressions we have omitted terms that include the Riemann tensor $\hat R_{ijk}{}^l$ associated with the connection $\Gammaflat^i_{jk}$, since these terms vanish for our case of a flat background.

The Hamiltonian constraint takes the form
\begin{eqnarray} \label{Ham}
{\mathcal H} & \equiv & \frac{2}{3} K^2 - \bar A_{ij} \bar A^{ij} + e^{- 4 \phi} ( \bar R - 8 \bar D^i \phi \bar D_i \phi - 8 \bar D^2 \phi) \nonumber \\
& & - 16 \pi \rho \nonumber \\
& = & 0,
\end{eqnarray}
while the momentum constraints can be written as
\begin{eqnarray}
{\mathcal M}^i & \equiv & e^{-4\phi} \Big( 
\frac{1}{\sqrt{\bar \gamma}} \Dflat_j(\sqrt{\bar \gamma} \bar A^{ij})
+ 6 \bar A^{ij} \partial_j \phi \nonumber \\ 
& & ~~~~~ - \frac{2}{3} \bar \gamma^{ij} \partial_j K + \bar A^{jk} \Delta \Gamma^i_{jk} 
\Big) - 8 \pi S^i
\nonumber \\
& = & 0.
\end{eqnarray}
(see equations (16) and (17) in \cite{Bro09}).

We note that when $\bar \gamma = 1$ and $\Gammaflat^i_{jk} = 0$, which is suitable for Cartesian coordinates, the above equations reduce to the traditional BSSN equations.   In the following, however, we will evaluate these equations in spherical polar coordinates.

Before the above equations can be integrated, we have to specify coordinate conditions for the lapse $\alpha$ and the shift $\beta^i$.  Unless noted otherwise we will adopt a ``non-advective" version of what has become the ``standard gauge" in numerical relativity.  Specifically, we use the ``1+log" condition for the lapse \cite{BonMSS95} in the form
\begin{equation} \label{1+log}
\partial_t \alpha = - 2 \alpha K,
\end{equation}
and the ``Gamma-driver" condition for the shift \cite{AlcBDKPST03} in the form
\begin{subequations} \label{gammadriver}
\begin{eqnarray}
\partial_t \beta^i & = & B^i \\
\partial_t B^i & = & \frac{3}{4} \partial_t \bar \Lambda^i.
\end{eqnarray}
\end{subequations}
(compare \cite{MonC12}).  These (or similar) conditions play a key role in the ``moving-puncture" approach to handling black hole singularities in numerical simulations.

\subsection{Implementation in spherical polar coordinates}
\label{sec:implementation}

We now focus on spherical polar coordinates, and will assume that the $\Gammaflat^i_{jk}$ are associated with the flat metric in spherical polar coordinates $r$,  $\theta$, and  $\phi$,
\begin{equation} \label{flatmetric}
\metricflat_{ij} = \eta_{ij} = \left(
\begin{array}{ccc}
1 & 0 & 0 \\
0 & r^2 & 0 \\
0 & 0 & r^2 \sin^2 \theta 
\end{array}
\right).
\end{equation}
Accordingly, the only non-vanishing components of the background connection are
\begin{equation} \label{flatconnection}
\begin{array}{rclrcl}
\Gammaflat^r_{\theta\theta} & = & -r  ~~~~ & ~~~~
\Gammaflat^r_{\phi\phi} & = & -r \sin^2 \theta  \\ 
\Gammaflat^\theta_{\phi\phi} & = & -\sin \theta \cos \theta ~~~~ & ~~~~ 
\Gammaflat^\theta_{r\theta} & = & r^{-1}  \\
\Gammaflat^\phi_{r\phi} & = & r^{-1}  ~~~~ & ~~~~
\Gammaflat^\phi_{\phi\theta} & = & \cot \theta.
\end{array}
\end{equation}

When implementing the above equations in spherical polar coordinates,
care has to be taken that coordinate singularities do not spoil the
numerical simulation.   These singularities appear both at the origin,
where $r = 0$, and on the axis where $\sin \theta = 0$.   Even for a
simple scalar wave, appearances of inverse factors of $r$ and $\sin
\theta$ in the Laplace operator can pose a challenge for a numerical
implementation.  In Section \ref{sec:numerics} below we discuss a PIRK method (see also \cite{CorCDI12,MonC12}) that handles these singularities very effectively.  

An additional challenge in general relativity is that these inverse factors of $r$ and $\sin \theta$ appear through the dynamical variables themselves.  Components of the spatial metric, for example, scale with powers of $r$ and $\sin \theta$, the inverse metric then scales with inverse powers of these quantities, and numerical error affecting these terms may easily spoil the numerical evolution.  It is therefore important to treat these appearances of $r$ and $\sin \theta$ analytically.  We therefore factor out suitable powers of $r$ and $\sin \theta$ from components of all tensorial objects.\footnote{In an alternative approach, one could represent the metric in an orthonormal frame, so that the correct powers of $r$ and $\sin\theta$ are absorbed in the unit vectors.}

We start by writing the conformally related metric $\bar \gamma_{ij}$ as the sum of the flat background metric $\metricflat_{ij}$ and a correction $\epsilon_{ij}$ (which is not assumed to be small),
\begin{equation} \label{metric}
\bar \gamma_{ij} = \metricflat_{ij} + \epsilon_{ij}.
\end{equation}
The flat metric $\metricflat_{ij}$ is given by eq.~(\ref{flatmetric}), and we write the correction
$\epsilon_{ij}$ in the form
\begin{equation}
\epsilon_{ij} = 
\left( \begin{array}{ccc}
h_{rr} 				&  r h_{r\theta}    			& r \sin \theta h_{r\phi} \\
r h_{r\theta} 			&  r^2 h_{\theta\theta} 		& r^2 \sin \theta h_{\theta \phi} \\
r \sin \theta h_{r\phi}  	& r^2 \sin \theta h_{\theta \phi} & r^2 \sin^2 \theta h_{\phi\phi}
\end{array}
\right).
\end{equation}
We similarly rescale the extrinsic curvature $\bar A_{ij}$ as
\begin{equation} \label{Acap}
\bar A_{ij} = 
\left( \begin{array}{ccc}
a_{rr} 				&  r a_{r\theta}    			& r \sin \theta a_{r\phi} \\
r a_{r\theta} 			&  r^2 a_{\theta\theta} 		& r^2 \sin \theta a_{\theta \phi} \\
r \sin \theta a_{r\phi}  	& r^2 \sin \theta a_{\theta \phi} & r^2 \sin^2 \theta a_{\phi\phi}
\end{array}
\right),
\end{equation}
and the connection vector $\bar \Lambda^i$ as 
\begin{equation} \label{lambda}
\bar \Lambda^i = 
\left( \begin{array}{c}
\lambda^r \\
\lambda^\theta/r \\
\lambda^\phi / (r \sin \theta) 
\end{array}
\right).
\end{equation}
We treat the shift $\beta^i$ and $B^i$ similarly, and finally  rewrite the evolution equations (\ref{evolution}) for the coefficients $h_{ij}$, $a_{ij}$ and $\lambda^i$ etc.

We can compute the connection coefficients (\ref{deltagamma}) from
\begin{equation} \label{connection}
\Delta \Gamma^i_{jk} = \frac{1}{2} \bar \gamma^{il} \left(
\Dflat_j \bar \gamma_{kl} + \Dflat_k \bar \gamma_{jl} - \Dflat_l \bar \gamma_{jk} \right).
\end{equation}
Since $\Dflat_i \metricflat_{jk} = 0$ we can compute the derivatives of the spatial metric
\begin{equation}
\Dflat_i \bar \gamma_{jk} = \Dflat_i \epsilon_{jk}
\end{equation}
in terms of the coefficients $h_{ij}$.  Direct calculation using the flat connection (\ref{flatconnection}) yields
\begin{equation} \label{metric_derivs}
\begin{array}{rcl}
\Dflat_r \bar \gamma_{rr} 			& = & h_{rr,r} \\
\Dflat_r \bar \gamma_{r\theta} 		& = & r h_{r\theta,r} \\
\Dflat_r \bar \gamma_{r\phi} 		& = & r \sin \theta h_{r\phi,r} \\
\Dflat_r \bar \gamma_{\theta\theta} 	& = & r^2 h_{\theta\theta,r} \\
\Dflat_r \bar \gamma_{\theta\phi} 	& = & r^2 \sin \theta h_{\theta\phi,r} \\
\Dflat_r \bar \gamma_{\phi\phi} 		& = & r^2 \sin^2 \theta h_{\phi\phi,r} \\ \\

\Dflat_\theta \bar \gamma_{rr} 			& = & h_{rr,\theta} - 2 h_{r\theta} \\
\Dflat_\theta \bar \gamma_{r\theta} 		& = & r (h_{r\theta,\theta} + h_{rr} - h_{\theta\theta} ) \\
\Dflat_\theta \bar \gamma_{r\phi} 		& = & r \sin \theta (h_{r\phi,\theta} - h_{\theta\phi}) \\
\Dflat_\theta \bar \gamma_{\theta\theta} 	& = & r^2 (h_{\theta\theta,\theta} + 2 h_{r\theta}) \\
\Dflat_\theta \bar \gamma_{\theta\phi} 	& = & r^2 \sin \theta (h_{\theta\phi,\theta} + h_{r\phi}) \\
\Dflat_\theta \bar \gamma_{\phi\phi} 		& = & r^2 \sin^2 \theta h_{\phi\phi,\theta} \\ \\

\Dflat_\phi \bar \gamma_{rr} 			& = & h_{rr,\phi} - 2 \sin \theta h_{r\phi}\\
\Dflat_\phi \bar \gamma_{r\theta} 		& = & r (h_{r\theta,\phi} - \cos \theta h_{r\phi} - 
	\sin \theta h_{\theta\phi}) \\
\Dflat_\phi \bar \gamma_{r\phi} 		& = & r \sin \theta (h_{r\phi,\phi} + \sin \theta h_{rr} 
	+ \cos \theta h_{r\theta} - \sin \theta h_{\phi\phi}) \\
\Dflat_\phi \bar \gamma_{\theta\theta} 	& = & r^2 (h_{\theta\theta,\phi} - 2 \cos \theta h_{\theta \phi}) \\
\Dflat_\phi \bar \gamma_{\theta\phi} 	& = & r^2 \sin \theta (h_{\theta\phi,\phi} + \sin \theta h_{r\theta} 
	+ \cos \theta h_{\theta\theta} - \cos \theta h_{\phi\phi}) \\
\Dflat_\phi \bar \gamma_{\phi\phi} 		& = & r^2 \sin^2 \theta (h_{\phi\phi,\phi} 
	+ 2 \sin \theta h_{r\phi}  + 2 \cos \theta h_{\theta\phi})
\end{array}
\end{equation}
The (flat) covariant derivative of the connection vector $\bar \Lambda^i$ can similarly be expressed in terms of the $\lambda^i$ as
\begin{equation}
\begin{array}{rcl}
\Dflat_r \bar \Lambda^r & = & \partial_r \lambda^r \\
\Dflat_\theta \bar \Lambda^r & = & \partial_\theta \lambda^r - \lambda^\theta \\
\Dflat_\phi \bar \Lambda^r & = & \partial_\phi \lambda^r - \sin \theta \lambda^\phi \\ \\ 
\Dflat_r \bar \Lambda^\theta & = & \displaystyle \frac{1}{r} \partial_r \lambda^\theta  \\[3mm]
\Dflat_\theta \bar \Lambda^\theta & = &\displaystyle \frac{1}{r} \left( 
	\partial_\theta \lambda^\theta + \lambda^r \right) \\[3mm]
\Dflat_\phi \bar \Lambda^\theta & = & \displaystyle \frac{1}{r} \left( 
	\partial_\phi \lambda^\theta - \cos \theta \lambda^\phi \right)  \\ \\[3mm]
\Dflat_r \bar \Lambda^\phi & = & \displaystyle \frac{1}{r \sin \theta} \partial_r \lambda^\phi  \\[3mm]
\Dflat_\theta \bar \Lambda^\phi & = &\displaystyle \frac{1}{r\sin\theta}  
	\partial_\theta \lambda^\phi  \\[3mm]
\Dflat_\phi \bar \Lambda^\phi & = & \displaystyle \frac{1}{r \sin\theta} \left( 
	\partial_\phi \lambda^\phi + \sin \theta \lambda^r + \cos \theta \lambda^\theta \right) 
\end{array}
\end{equation}
Using the above expressions, we can compute the Ricci tensor (\ref{ricci}) as follows.  In the first 
term on the right-hand side of (\ref{ricci}) we write the second covariant derivative of $\bar \gamma_{ij}$ as a sum of first partial derivatives of the quantities $\Dflat_i \bar \gamma_{ij}$ and (flat) connection terms multiplying the $\Dflat_i \bar \gamma_{ij}$,
\begin{eqnarray}
\Dflat_k \Dflat_l \bar \gamma_{ij} & = & \partial_k (\Dflat_l \bar \gamma_{ij})  \\
& & - (\Dflat_m \bar \gamma_{ij}) \Gammaflat^m_{lk} 
- (\Dflat_l \bar \gamma_{mj}) \Gammaflat^m_{ik} 
- (\Dflat_l \bar \gamma_{im}) \Gammaflat^m_{jk}.  \nonumber
\end{eqnarray} 
We then insert the expressions (\ref{metric_derivs}) into the first term on the right-hand side and evaluate all derivatives explicitly, so that these terms can be written in terms of second partial derivatives of the coefficients $h_{ij}$.  Once this step has been completed, we add those remaining terms for which the flat background connection (\ref{flatconnection}) is nonzero.

The resulting equations are rather cumbersome, and it is easy to introduce typos in the numerical code.  The numerical examples of Section \ref{sec:numerical_examples} are excellent tests of the code.  In Appendix \ref{appendixA} we describe another analytical test that we have found very useful to check our implementation of curvature quantities.

As a final comment we note that the condition (\ref{dgammadt}) determines the time evolution of the determinant $\bar \gamma$ of the conformally related metric, but not its initial value.  The latter can be chosen freely in this scheme, in particular it does not need to be chosen equal to that of the background metric $\hat \gamma$ (unlike in the original BSSN formulation).  For some of our numerical simulations, however, in particular for the rotating star simulations of Section \ref{sec:rot_star}, we found that rescaling the conformally related metric so that its determinant becomes $\hat \gamma = r^4 \sin^2 \theta$ improved the stability of the simulation, so that it required a smaller coefficient $\eta$ in the Kreiss-Oliger dissipation term (\ref{KO}) below.

\section{Numerical Implementation}
\label{sec:numerics}

\subsection{PIRK methods}
\label{sec:PIRK}

The origin of the numerical instabilities in curvilinear coordinate
systems are related to the presence of stiff source terms in the
equations, e.g.~factors of $1/r^2$ or $1/\sin^2(\theta)$ that become
arbitrary large close to the origin or the axis.  In the following we
will refer to these terms as ``singular terms''.   PIRK methods
evolve all other, i.e.~regular, terms in the evolution equations
explicitly, and then use these updated values to evolve the singular
terms implicitly.  This strategy implies that the computational costs
of PIRK methods are comparable to those of explicit methods. The
resulting numerical scheme does not need any analytical or numerical
inversion, but is able to provide stable evolutions due to its
partially implicit component. We refer to~\cite{CorCD12} for a
detailed derivation of PIRK methods (up to third order), and limit our discussion here
to a simple description of the second-order PIRK method that is
implemented in our code.

Consider a system of partial differential equations
\begin{System}
u_t = \mathcal{L}_1 (u, v), \\
v_t = \mathcal{L}_2 (u) + \mathcal{L}_3 (u, v),
\label{e:system}
\end{System}
where $\mathcal{L}_1$, $\mathcal{L}_2$ and $\mathcal{L}_3$ are general
non-linear differential operators.  We will denote the corresponding
discretized operators by $L_1$, $L_2$ and $L_3$, respectively.  We
will further assume that $L_1$ and $L_3$ contain only regular terms,
and hence will update these terms explicitly, whereas the $L_2$ operator 
contains the singular terms and will therefore be treated
partially implicitly.  Note that $L_2$ is assumed to depend on $u$ only.  In the case of the BSSN equations this holds for almost all variables; the one exception can be treated as discussed in the paragraph below equation (\ref{L3Lambda}) in Appendix \ref{appendixB}, where we provide the exact form of the source terms.

In our second-order PIRK scheme we update the variables $u$ and $v$ from an old timestep $n$ to a new timestep $n+1$ in two stages.  In each of these two stages, we first evolve the variable $u$ explicitly, and then evolve the variable $v$ taking into account the updated values of $u$ for the evaluation of the singular $L_2$ operator.  For the system of equations (\ref{e:system}), the first stage 
\begin{System}
	u^{(1)} = u^n + \Delta t \, L_1 (u^n, v^n),  \\
	v^{(1)} = v^n + \Delta t \left[\frac{1}{2} L_2(u^n) +
          \frac{1}{2} L_2(u^{(1)}) + L_3(u^n, v^n) \right],
\end{System}
is followed by the second stage
\begin{System}
	u^{n+1}  = \frac{1}{2} \left[ u^n + u^{(1)} 
+ \Delta t \, L_1 (u^{(1)}, v^{(1)}) \right], \\
	v^{n+1}  =   v^n + \frac{\Delta t}{2} \left[ 
L_2(u^n) + L_2(u^{n+1}) \right.  \\ 
\left. \hspace{2.5cm} +  L_3(u^n, v^n) + L_3 (u^{(1)}, v^{(1)}) \right].
\end{System}

In the first stage, $u$ is evolved explicitly; the updated value
$u^{(1)}$ is used in the evaluation of the $L_2$ operator for the
computation of $v^{(1)}$.  In the second stage, $u$ is again evolved
explicitly, and the updated value $u^{n+1}$ is used in the evaluation
of the $L_2$ operator for the computation of the updated values $v^{n+1}$.

Our PIRK method is stable as long as the timestep is limited by a
Courant condition; see eq.~(\ref{Courant}) below.

We include all singular terms appearing in the sources of the
equations in the $L_2$ operator. Firstly, the conformal metric
components, $h_{ij}$, the conformal factor, $\phi$, the lapse function,
$\alpha$, and the shift vector, $\beta^i$, are evolved explicitly (as
$u$ is evolved in the previous PIRK scheme); secondly, the traceless
part of the extrinsic curvature, $a_{ij}$, and the trace of the
extrinsic curvature, $K$, are evolved partially implicitly, using
updated values of $\alpha$, $\beta^i$, $\phi$ and $h_{ij}$; then, the
$\lambda^{i}$ are evolved partially implicitly, using the updated
values of $\alpha$, $\beta^i$, $\phi$, $h_{ij}$, $a_{ij}$ and
$K$. Finally, $B^i$ is evolved partially implicitly, using the updated
values of the previous quantities. Lie derivative terms and matter
source terms are always included in the explicitly treated parts. In
Appendix~\ref{appendixB}, we give the exact form of the source terms
included in each operator.

\subsection{Numerical grid}

\begin{figure}[t]
\includegraphics[width=3.4in]{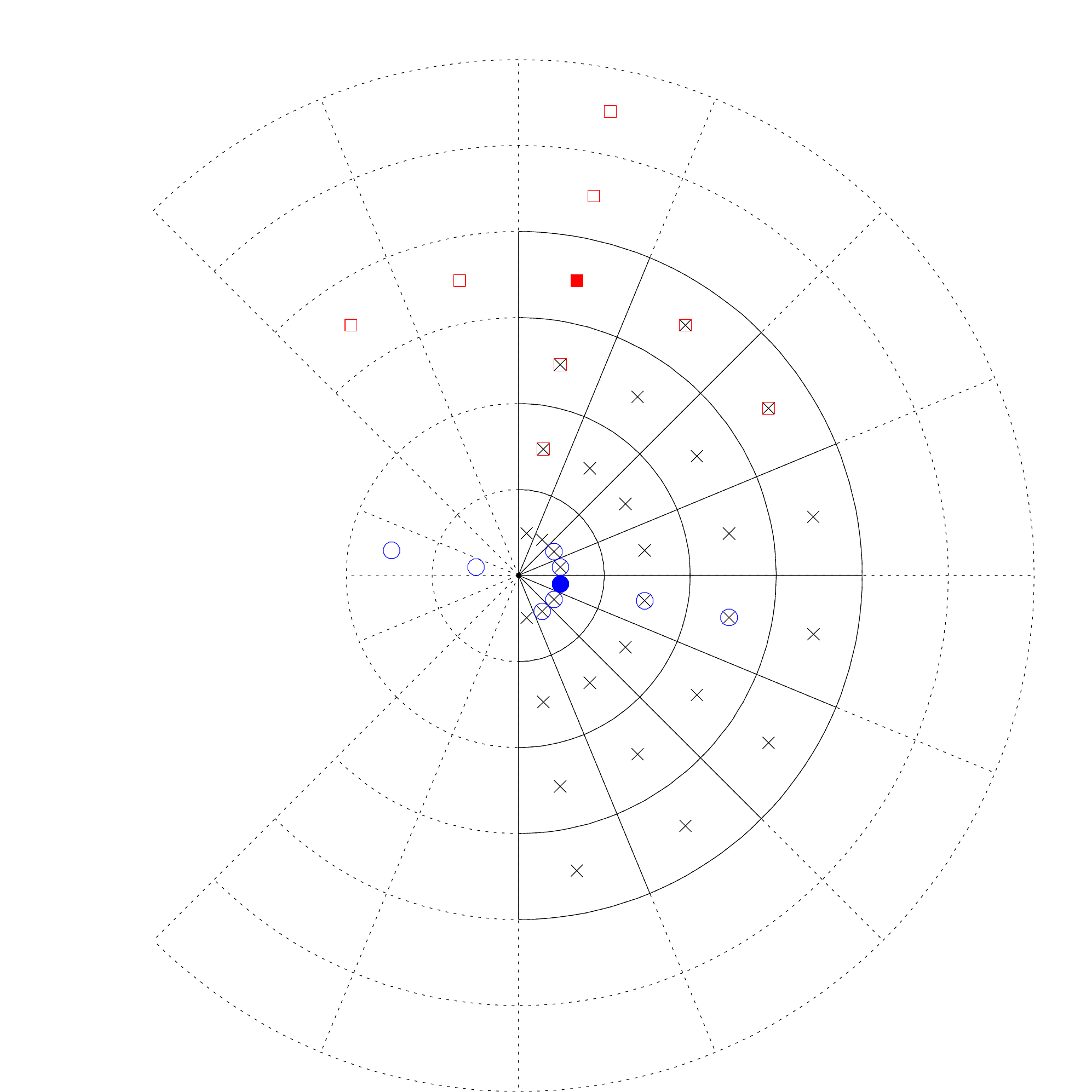}
\caption{A schematic representation of our cell-centered grid structure in spherical polar coordinates, for one fixed value of $\phi$.  Grid points, marked by the crosses, are placed at the center of grid cells, so that no grid point ends up at the center ($r=0$) or on the axes  ($\sin \theta = 0$ or $\sin \theta = \pi$).  Our interior grid, bordered by solid lines in the figure, covers the region $0 \leq r \leq r_{\rm max}$ and $0 \leq \theta \leq \pi$ (as well as $0 \leq \phi \leq 2 \pi$).  As suggested by the two highlighted stencils, our fourth-order differencing scheme requires two levels of ghost zones outside of the interior grid, indicated by the dotted lines.}   
\label{Fig1}
\end{figure}

We adopt a centered, fourth-order finite differencing representation
of most spatial derivatives.  For each grid point, the
finite-differencing stencil therefore involves the two nearest
neighbors in each direction (see Fig.~\ref{Fig1}).  An exception from
our fourth-order differencing are advective derivatives along the
shift, for which we use a second-order (one-sided) upwind scheme.
Because of the second-order time evolution, and the second-order
advective terms, our scheme is overall second-order accurate, even
though for some cases with vanishing shift we have found that the
error appears to be dominated by the fourth-order terms. 

We adopt a cell-centered grid, as shown schematically in Fig.~\ref{Fig1}.  Specifically, we divide the physical domain covered by our grid, $0 < r < r_{\rm max}$, $0 < \theta < \pi$ and $0 < \phi < 2 \pi$ into $N_r \times N_\theta \times N_\phi$ cells with uniform coordinate size
\begin{equation}
\Delta r = r_{\rm max}/ N_r,~~~~
\Delta \theta = \pi/ N_\theta,~~~~
\Delta \phi = 2 \pi/ N_\phi.
\end{equation}
Because of our fourth-order finite differencing scheme we need to pad the interior grid with two layers of ghost zones.  Except at the outer boundary, each ghost zone corresponds to some other zone in the interior of the grid (with some other value of $\theta$ and $\phi$), so that these ghosts zones can be filled by copying the corresponding values from interior grid points.  

As a concrete example, consider a grid point with angular coordinates $\theta$ and $\phi$, say, in the innermost radial zone (highlighted by a (blue) filled circle in Fig.~\ref{Fig1}).  Evaluating the partial derivative with respect to $r$ at this point requires two grid points that, formally, have negative radii.  We can fill these two required ghost points by finding the corresponding points in the interior of the grid, which have angular coordinates $\pi - \theta$ and $\phi + \pi$.  Similarly, evaluating a derivative with respect to $\theta$ for a point with angular coordinates $(\theta,\phi)$ next to the axis (see the (red) filled square in Fig.~\ref{Fig1}) requires ghost points that can be filled by finding the corresponding grid points with azimuthal angle $\phi + \pi$ in the interior of the grid.

\begin{table}[t]
\begin{tabular}{|c|c|c|c|}
\hline
 & & center & axis \\
 \hline
\multirow{3}{0.5in}{vectors} & $r$ & - & + \\
& $\theta$ & + & - \\
& $\phi$  & - & - \\
\hline
\multirow{6}{0.5in}{tensors} & $rr$ & + & + \\
& $r\theta$ & - & - \\
& $r\phi$ & + & - \\
& $\theta\theta$ & + & + \\
& $\theta\phi$ & - & + \\
& $\phi\phi$ & + & + \\
\hline
\end{tabular}
\caption{Parity conditions for components of vectors and tensors as implemented in our coordinate-based code.  Components of vectors and tensors have to be multiplied with the corresponding sign when they are copied into ghost zones at the center or the axis.} 
\label{Tab1}
\end{table}

For scalar functions the corresponding function values can be copied immediately, but for components of vectors or tensors, expressed in spherical polar coordinates, a possible relative sign has to be taken into account.  Essentially, this occurs because, in spherical polar coordinates, the unit vectors may point into the opposite physical direction when we identify a ghost zone with an interior point, i.e.~when we go from $(\theta,\phi)$ to $(\pi - \theta, \phi + \pi)$ or $(\theta, \phi + \pi)$.   We list these relative sign changes, as implemented in our coordinate-based code, in Table \ref{Tab1}.

We also require two sets of two ghost zones for $\phi$, which can be filled directly using periodicity.

At the outer boundary we also require two ghost zones, as suggested by the (red) squared stencil in Fig.~\ref{Fig1}.  We impose a Sommerfeld boundary condition, which is an approximate implementation of an outgoing wave boundary condition, to fill these ghost zones.  In our coordinate-based code we implement this condition by tracing an outgoing radial characteristic from each of the outer boundary grid points back to the previous time level.  We then interpolate the corresponding function to the intersection of that characteristic and the previous time level, and copy that interpolated value, multiplied by a suitable fall-off in $r$, into the boundary grid point.  We assume a fall-off with $r^{-1}$ for all metric variables (i.e.~$h_{ij}$, $a_{ij}$, $\phi$ and $K$) as well as the lapse $\alpha$, but a $r^{-2}$ fall-off for the shift $\beta^i$ as well as $\lambda^i$.  

The PIRK method of Section \ref{sec:PIRK} is stable as long as the
time step $\Delta t$ is limited by a Courant-Friedrichs-Lewy
condition.  In order to evaluate this condition we first find the
smallest coordinate distance $\Delta_{\rm min}$ between any two
grid-points in our cell-centered, spherical polar grid. This minimum
distance is approximately
\begin{equation}
\Delta_{\rm min} = \min(\Delta r, (\Delta r/2) \Delta \theta, (\Delta r/2) \sin (\Delta \theta/2) \Delta \phi).
\end{equation}
We then set
\begin{equation} \label{Courant}
\Delta t = C \Delta_{\rm min},
\end{equation}
where we have chosen a Courant factor $C = 0.4$ for all simulations in this paper.  It is a well-known disadvantage of spherical polar coordinates that the accumulation of gridpoints in the vicinity of the origin leads to a very severe limit on the timestep.  We will discuss this issue in greater detail in Section \ref{sec:discussion}.

We use Kreiss-Oliger~\cite{KreO73} dissipation to suppress the appearance of
high frequency noise at late times.    Specifically, we add a term of the form
\begin{equation} \label{KO}
f_{\rm KO} = \frac{\eta}{16 \Delta t} \left( 
	(\Delta r)^4 \frac{\partial^4 f}{\partial r^4} +
	(\Delta \theta)^4 \frac{\partial^4 f}{\partial \theta^4} +
	(\Delta \phi)^4 \frac{\partial^4 f}{\partial \phi^4} \right)
\end{equation}
to the right hand side of the evolution equation for each dynamical variable $f$.  Here $\eta$ is a dimensionless coefficient which we have chosen between 0 (for some of our short-time evolutions) and 0.001 for the rotating neutron star simulation in Sect.~\ref{sec:rot_star}.

\section{Numerical Examples}
\label{sec:numerical_examples}

\subsection{Weak gravitational waves}

As a first test of our codes we consider small-amplitude gravitational waves on a flat Minkowski background.  Following Teukolsky \cite{Teu82} we construct an analytical, linear solution for quadrupolar ($\ell = 2$) waves from a function 
\begin{equation}
F(r,t) = A (r \pm t) e^{-(r \pm t)^2/\lambda^2},
\end{equation}
where the constant $A$ is related to the amplitude of the wave and $\lambda$ to its wavelength (see also Section 9.1 in \cite{BauS10}).  We set $\lambda = 1$, by which all length scales become dimensionless.  We will consider axisymmetric ($m=0$) and non-axisymmetric ($m=2$) modes separately.

\subsubsection{Axisymmetric waves}

\begin{figure}[t]
\includegraphics[width=3.4in]{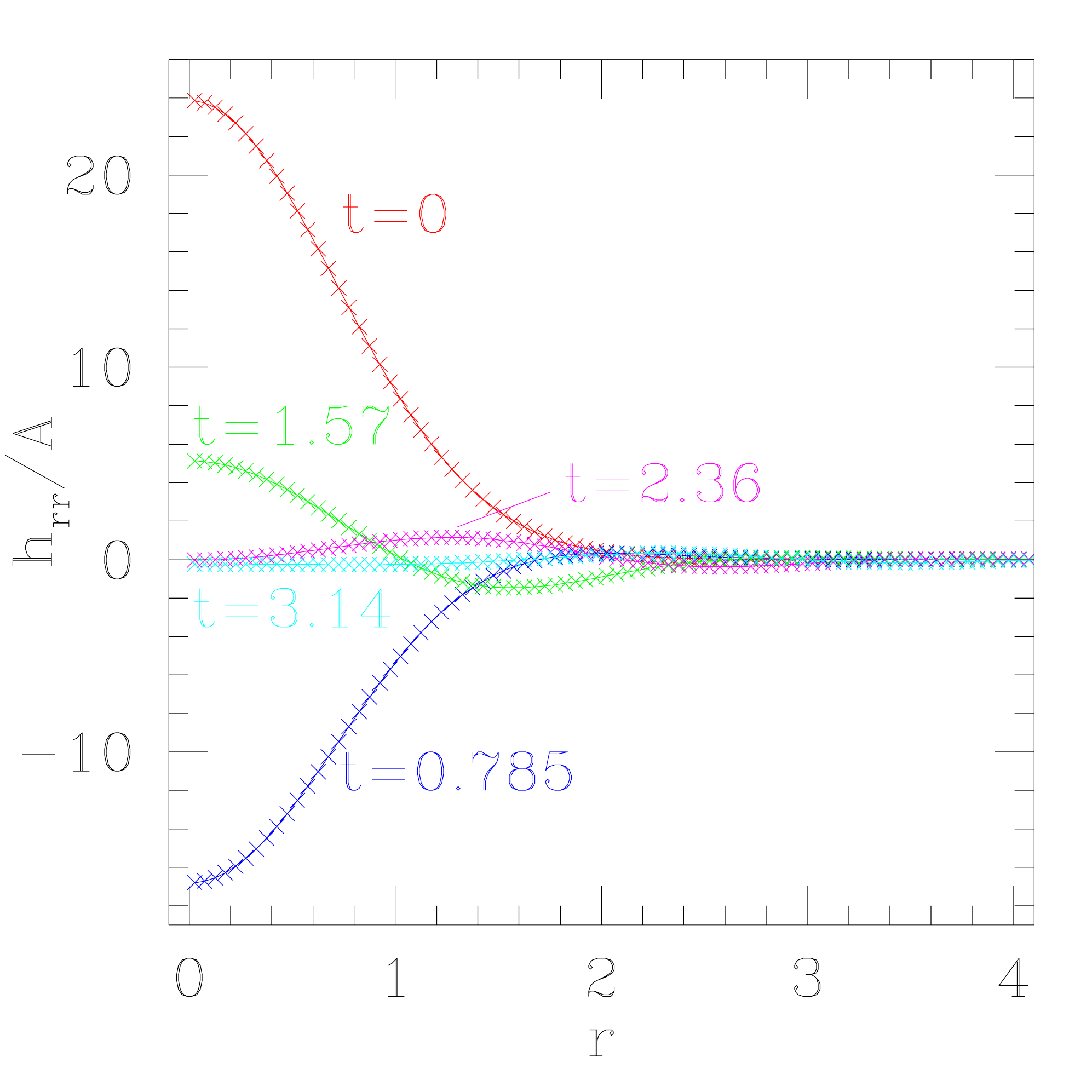}
\caption{Snapshots of the metric coefficient $h_{rr}$ for an axisymmetric $m=0$ small-amplitude gravitational wave at different instances of time.  For this simulation we used a grid of size $(160,40,2)$ and imposed the outer boundary at $r_{\rm max} = 8.0$.  We show data as a function of $r$ in the (arbitrary) direction $\theta = 1.61$ and $\phi = 4.71$.  Differences between the numerical results (marked by crosses) and the analytical solution (solid lines) are smaller than the width of the lines in this graph.}
\label{Fig2}
\end{figure}

\begin{figure}[t]
\includegraphics[width=3.4in]{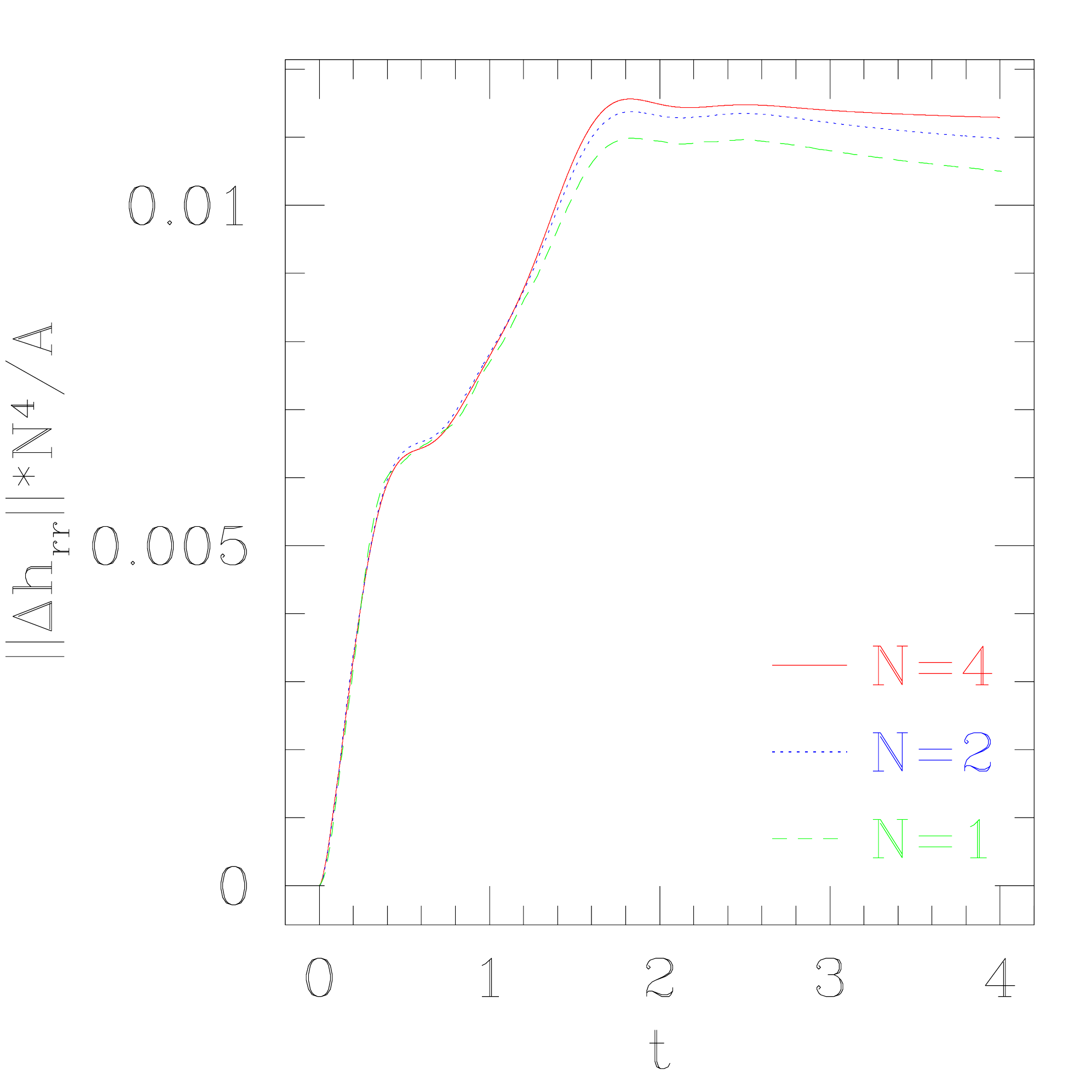}
\caption{The norm of the error in the quantity $h_{rr}$ as a function of time for a small-amplitude, axisymmetric gravitational wave.  We show results for simulations with a grid of size $(40N,10N,2)$, for $N=1$, $N=2$ and $N=4$, with the outer boundary imposed at $r = 8.0$.  At these early times, the error appears to be dominated by the fourth-order differencing of the spatial derivatives.}
\label{Fig3}
\end{figure}

We first consider axisymmetric $m=0$ waves.  Since these solutions are independent of the coordinate $\phi$, we may choose $N_\phi$ as small as possible (which is $N_\phi = 2$ in our code) without loss of accuracy.   We also choose a small amplitude of $A = 10^{-7}$, so that deviations from the analytic solution, which is accurate only to linear order in $A$, are dominated by our finite-difference error, and not by terms that are higher-order in $A$.  

In the following we show results for a numerical grid with $(40N,10N,2)$ grid points, where $N=1$, $N=2$ or $N=4$, and imposing the outer boundary at $r = 8.0$.  For these simulations we 
used the 1+log lapse condition (\ref{1+log}), but chose a vanishing shift $\beta^i = 0$ instead of the Gamma-driver condition (\ref{gammadriver}).

In Fig.~\ref{Fig2} we show snapshots of the metric function $h_{rr}$ at different instances of time for our highest-resolution simulation with $N = 4$.  For each time, we include the numerical results as crosses, as well as the analytical solution as a solid line.  The differences between the numerical results and analytical solution are well below the resolution limit of this graph, so that the two cannot be distinguished in this Figure.

In Fig.~\ref{Fig3} we show a convergence test for these waves.  Specifically, we compute the $L^2$-norm of the difference between the analytical solution $h_{rr}$ and the analytical solution,
\begin{equation} \label{norm}
|| \Delta h_{rr} || = \frac{1}{V} \left( \int (h_{rr}^{\rm num} - h_{rr}^{\rm ana})^2 dV \right)^{1/2},
\end{equation}
where $V$ is the coordinate volume of the numerical grid.   In Fig.~\ref{Fig3} we show these norms as a function of time for $N=1$, $N=2$ and $N=4$.   The norms are rescaled with a factor $N^4$; the convergence of the resulting error curves indicates that, at these early times, the error is dominated by the fourth-order differencing of the spatial derivatives.  In spherical polar coordinates, the Courant condition (\ref{Courant}) limits the time step to such small values that the second-order errors associated with our PIRK method are smaller than the fourth-order error of our spatial derivatives (for vanishing shift).

\subsubsection{Nonaxisymmetric waves}

\begin{figure}[t]
\includegraphics[width=3.4in]{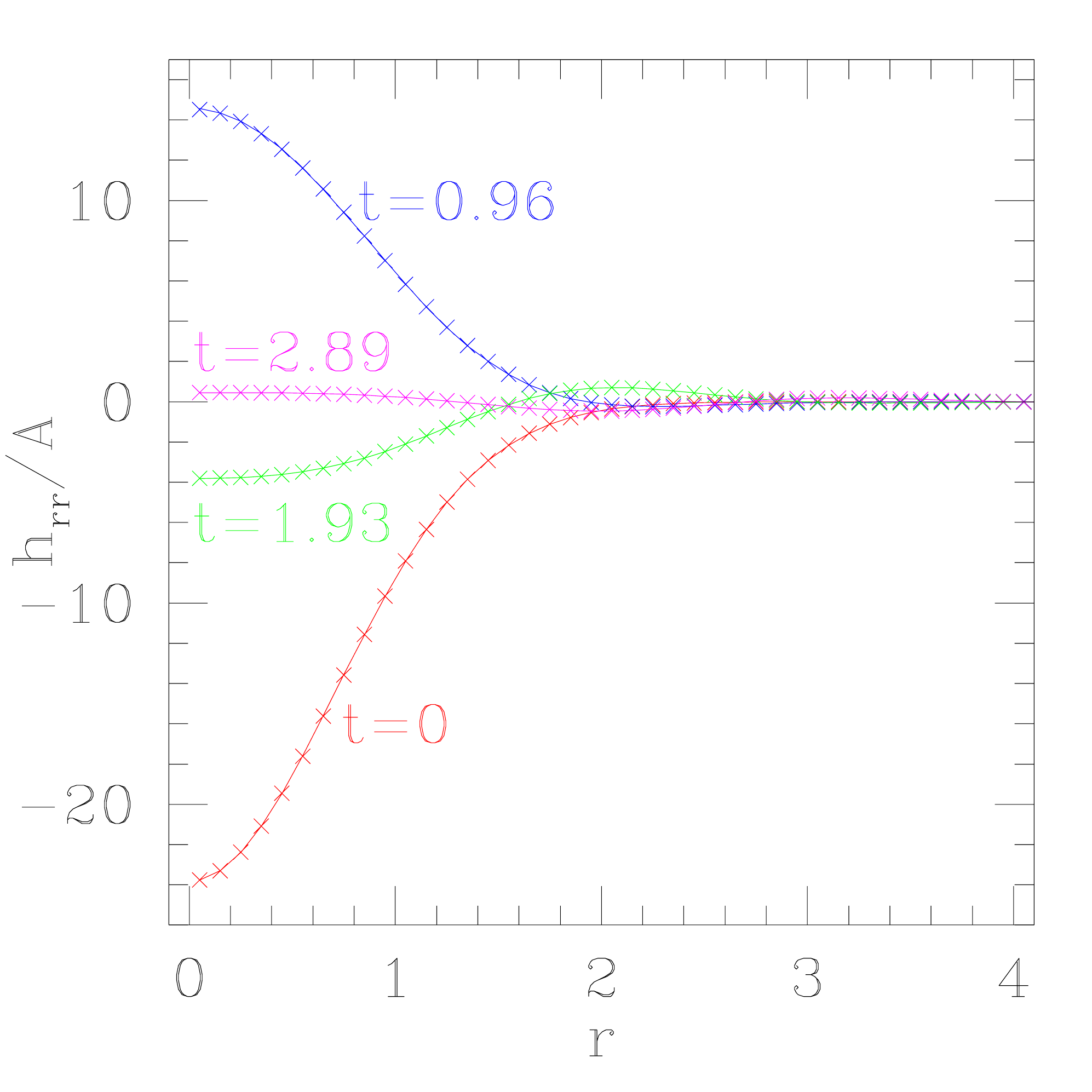}
\caption{Snapshots of the metric coefficient $h_{rr}$ for a non-axisymmetric $m=2$ small-amplitude gravitational wave at different instances of time.  For this simulation we used a grid of size $(40,32,64)$ and imposed the outer boundary at $r_{\rm max} = 4.0$; we show data as a function of $r$ in the direction $\theta = 1.62$ and $\phi = 3.19$.  Numerical results are marked by the crosses, while the analytical solution is shown as the solid line.}
\label{Fig4}
\end{figure}

Non-axisymmetric gravitational waves represent a rare example of an analytical, time-dependent, three-dimensional, albeit weak-field solution to the Einstein equations.  Clearly, this solution represents a stringent test for our code.\footnote{For a code in Cartesian coordinates, even a spherically symmetric solution represents a stringent test, because the symmetry is not reflected by the numerical grid.  In our code, however, numerical expressions simplify for spherical or axisymmetric solutions, so that they do not test every aspect of the code.}

In Fig.~\ref{Fig4} we show results for an $m=2$ wave, again for an amplitude $A = 10^{-7}$.  As in Fig.~\ref{Fig2}, we graph solutions for $h_{rr}$ as functions of $r$ at different instances of time.  Again, our numerical solution (marked by crosses) can hardly be distinguished from the analytical solution (shown as solid lines).

\subsection{Hydro-without-hydro}

As a test of strong-field, but regular solutions we consider spacetimes containing relativistic stars.  In general, this requires evolving the stellar matter self-consistently with the gravitational fields, for example by solving the equations of relativistic hydrodynamics.  Since this is beyond the scope of this paper, we here adopt the ``hydro-without-hydro'' approach suggested by \cite{BauHS99}.  In this approach, which can also be described as an ``inverse-Cowling approximation'', we leave the matter sources fixed, and evolve only the gravitational fields.  In this way, it is possible to assess the stability of a spacetime evolution code, and its capability of accurately evolving strong but regular gravitational fields in spacetimes with static matter, without having to worry about the hydrodynamical evolution. These simulations serve as both a testbed and a preliminary step towards fully relativistic hydrodynamical simulations of stars.  In this Section we consider static and uniformaly rotating stars separately.

\subsubsection{Spherical neutron stars}
\label{sec:TOV}

We first consider non-rotating relativistic stars, described by the Tolman-Oppenheimer-Volkoff (TOV) solution \cite{Tol39,OppV39}.  We focus on a polytropic TOV star with polytropic index $\Gamma=2$, and with a gravitational mass of about 85\% of the maximum-allowed mass.  For this model, the central density is about 40\% of that of the maximum mass model.  We evolved this star with the 1+log slicing condition for the lapse (\ref{1+log}), but kept the shift fixed to zero.  Because the spacetime is spherically symmetric, we could choose both $N_\theta$ and $N_\phi$ as small as possible ($N_\theta = N_\phi = 2$) without loss of accuracy.

Even for very modest grid resolutions in the radial direction (e.g.~$N_r = 40$, with the outer boundary imposed at four times the stellar radius), we found that the gravitational fields settle down into an equilibrium that is similar to the initial data.  After this initial transition, which is caused by the finite-difference error, the stellar surface as well as the outer boundaries (see \cite{BauHS99}), the solution remains stable.

\subsubsection{Rotating neutron stars}
\label{sec:rot_star}

\begin{figure}[t]
\includegraphics[width=3.4in]{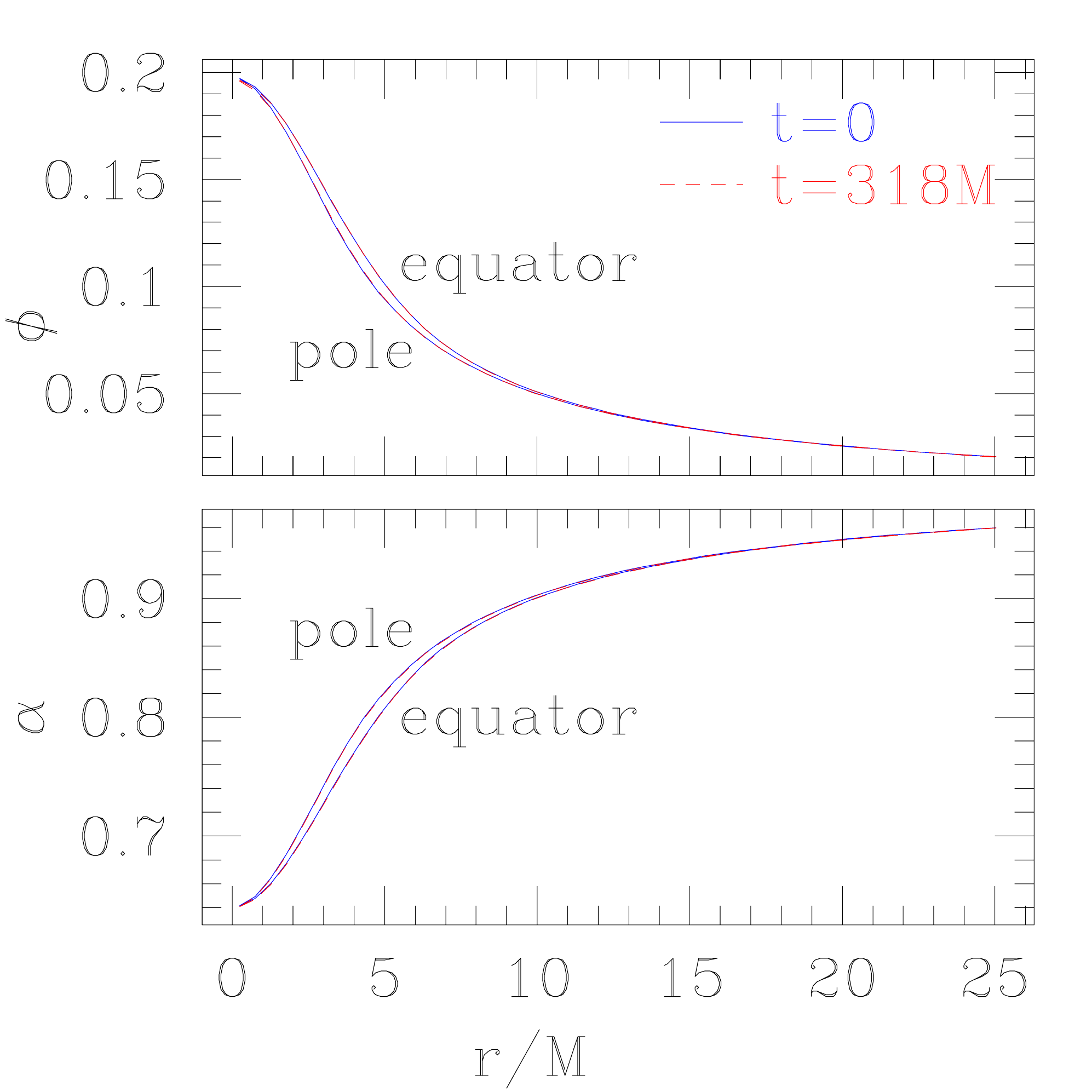}
\caption{Snapshots of the conformal exponent $\phi$ and the lapse $\alpha$ for a rapidly rotating star (see text for details). We show both functions both at the initial time, and at a late time $t = 318M$.  We also show both functions along rays in two different directions, one very close to the equator, the other pointing close to the pole.  Both profiles remain very similar to their initial data throughout the evolution.}
\label{Fig5}
\end{figure}

The evolution of the spacetime of a rapidly rotating relativistic star
is a more demanding test than the previous one, as it breaks spherical
symmetry and instead involves axisymmetric non-vacuum initial data in
the strong gravity regime. The initial data used for this test are the
numerical solution of a stationary and axisymmetric equilibrium model
of a rapidly and uniformly rotating relativistic star~\cite{BonGSM93},
which is computed using the Lorene code \cite{lorene}.

We consider a uniformly rotating star with the same $\Gamma=2$ polytropic equation of state as the non-rotating model of Sect.~\ref{sec:TOV}.  Our particular model has the same central rest-mass density as that non-rotating model, but rotates at $92\%$ of the allowed mass-shedding limit (for a star of that central density); expressed in terms of the gravitational mass $M$, the corresponding spin period is approximately 157 $M$. The ratio of the polar to
equatorial coordinate radii for this model is $0.7$.
For this simulation we adopted both the 1+log condition for the lapse (\ref{1+log}) and the
Gamma-driver condition for the shift (\ref{gammadriver}).

For this test we adopted a grid of size $(48,32,2)$, and imposed the outer boundary at $25.5M$, which equals four times the equatorial radius.  In Fig.~\ref{Fig5} we show the initial and late-time profiles of the conformal exponent $\phi$ and the lapse $\alpha$, both in a direction close to the equator and close to the axis.  Evidently, both functions remain very close to their initial values throughout the evolution, as they should.

\subsection{Schwarzschild}

In this Section we present results for two different simulations involving Schwarzschild black holes.
In Section \ref{sec:trumpet} we evolve a Schwarzschild black hole in a ``trumpet" geometry \cite{HanHPBO06,HanHOBGS06,HanHOBO08}, which, in the limit of infinite resolution, is a time-independent solution to the Einstein equations given our slicing conditions (\ref{1+log}).  In Section \ref{sec:wormhole} we adopt wormhole initial data and follow the coordinate transition to a trumpet geometry.

\subsubsection{Trumpet initial data}
\label{sec:trumpet}

\begin{figure}[t]
\includegraphics[width=3.4in]{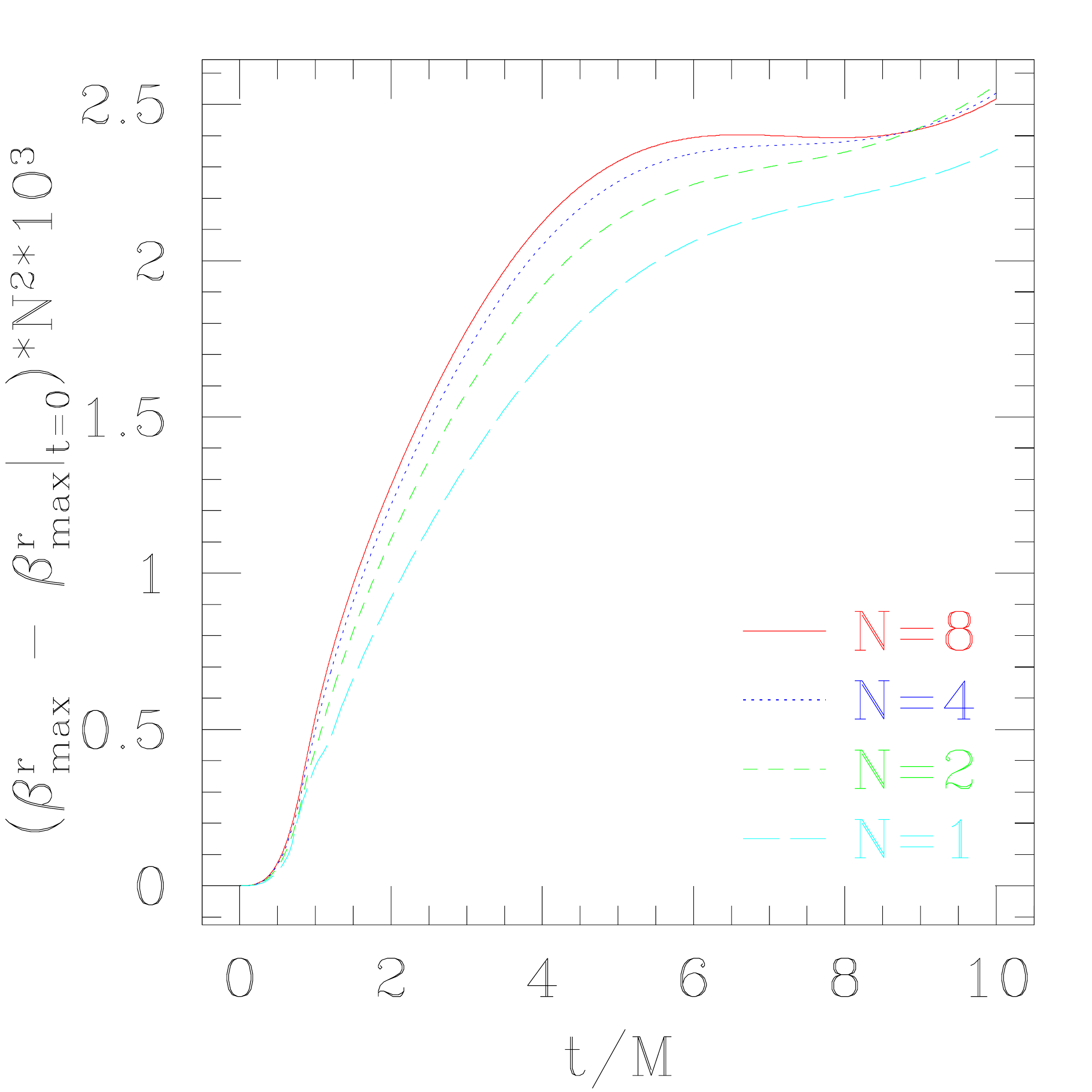}
\caption{The difference between the maximum of the radial component of the shift vector, $\beta^r_{\rm max}$, and its initial value $\beta^r_{\rm max}|_{\rm t=0}$, as a function of time.  For these simulations we used a grid of size $(160N,2,2)$ for $N=1$, 2, 4 and 8, and imposed the outer boundary at $r_{\rm max} = 16.0 M$.  We rescale all differences with $N^2$, so that the convergence of these lines demonstrates second-order convergence.}
\label{Fig6}
\end{figure}

Maximally sliced trumpet data \cite{HanHOBGS06} represent a time-independent slicing of the Schwarzschild spacetime that satisfies our slicing condition (\ref{1+log}).  The solution can be expressed analytically in isotropic coordinates, albeit only in parametrized form \cite{BauN07}.  In this Section we adopt these trumpet data as initial data, so that, in the continuum limit, the solution should remain independent of time.

For trumpet data the conformal factor diverges at $r=0$.  While, on our cell-centered grid, functions are never evaluated directly at the origin, derivatives in the neighborhood of the singularity at the origin are clearly affected by the singular behavior of the conformal factor.  However, the great virtue of the ``moving-puncture" gauge conditions (\ref{1+log}) and (\ref{gammadriver}) is that these errors only affect the neighborhood of the puncture, and do not spoil the evolution globally \cite{CamLMZ06,BakCCKM06a,HanHPBO06,Bro08}.  In the following we will demonstrate these properties in our code using spherical polar coordinates.

For the simulations presented in this section we adopted a numerical grid of size size $(160N,2,2)$ for $N=1$, 2, 4 and 8, with the outer boundary imposed at $r_{\rm max} = 16.0 M$.

In Fig.~\ref{Fig6} we show results for the maximum of the radial component $\beta^r$ of the shift vector as a function of time.  Specifically, we show the difference between these maximum values $\beta^r_{\rm max}$ and their initial values $\beta^r_{\rm max}|_{t=0}$.   Since our trumpet data represent a time-independent solution to the Einstein equations and our slicing and gauge conditions (\ref{1+log}) and (\ref{gammadriver}), these differences should converge to zero as the grid resolution is increased.  In Fig.~\ref{Fig6} we multiply the differences with $N^2$; the convergence of the resulting lines therefore demonstrates second-order convergence of the simulation.  Apparently the error in these simulations is dominated by the second-order advective terms.

We also note that the outer boundary introduces error terms that depend on both the grid resolution and the location of the outer boundary.   Since the latter does not decrease when we increase the grid resolution, the code converges more slowly in regions that have come into causal contact with the outer boundary.  We therefore include in Fig.~\ref{Fig6} only sufficiently early times, before the location of the shift's maximum is affected by the outer boundary.

\subsubsection{Wormhole initial data}
\label{sec:wormhole}

\begin{figure}[t]
\includegraphics[width=3.4in]{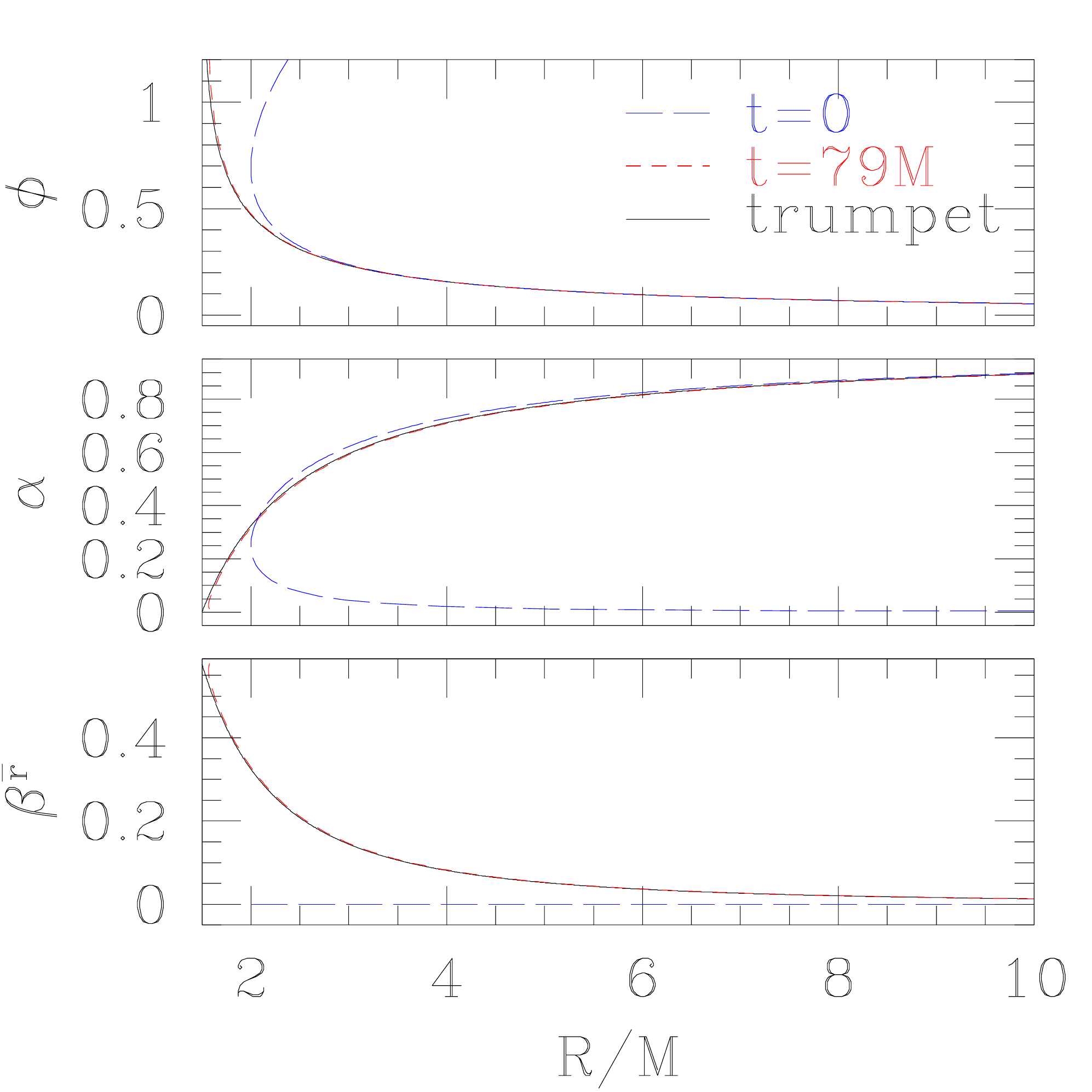}
\caption{Initial data and final profiles of the conformal exponent $\phi$, the lapse function $\alpha$, and the shift $\beta^{\hat r}$, showing the coordinate transition from wormhole initial data to time-independent trumpet data.  The (blue) long-dashed lines represent the initial data at $t=0$, the (red) dashed lines show our numerical results at time $t =79M$, and the (black) solid lines show the analytical trumpet solution \cite{BauN07}.  The initial data appear double-valued because we graph this functions as a function of the areal radius $R$ (see text for details).  For these simulations we adopted a grid size $(10240,2,2)$ with the outer boundary imposed at $r = 256M$. (In these graphs we did not include the innermost two grid points, which are affected by the singular behavior of the puncture.)}
\label{Fig7}
\end{figure}

\begin{figure}[t]
\includegraphics[width=3.4in]{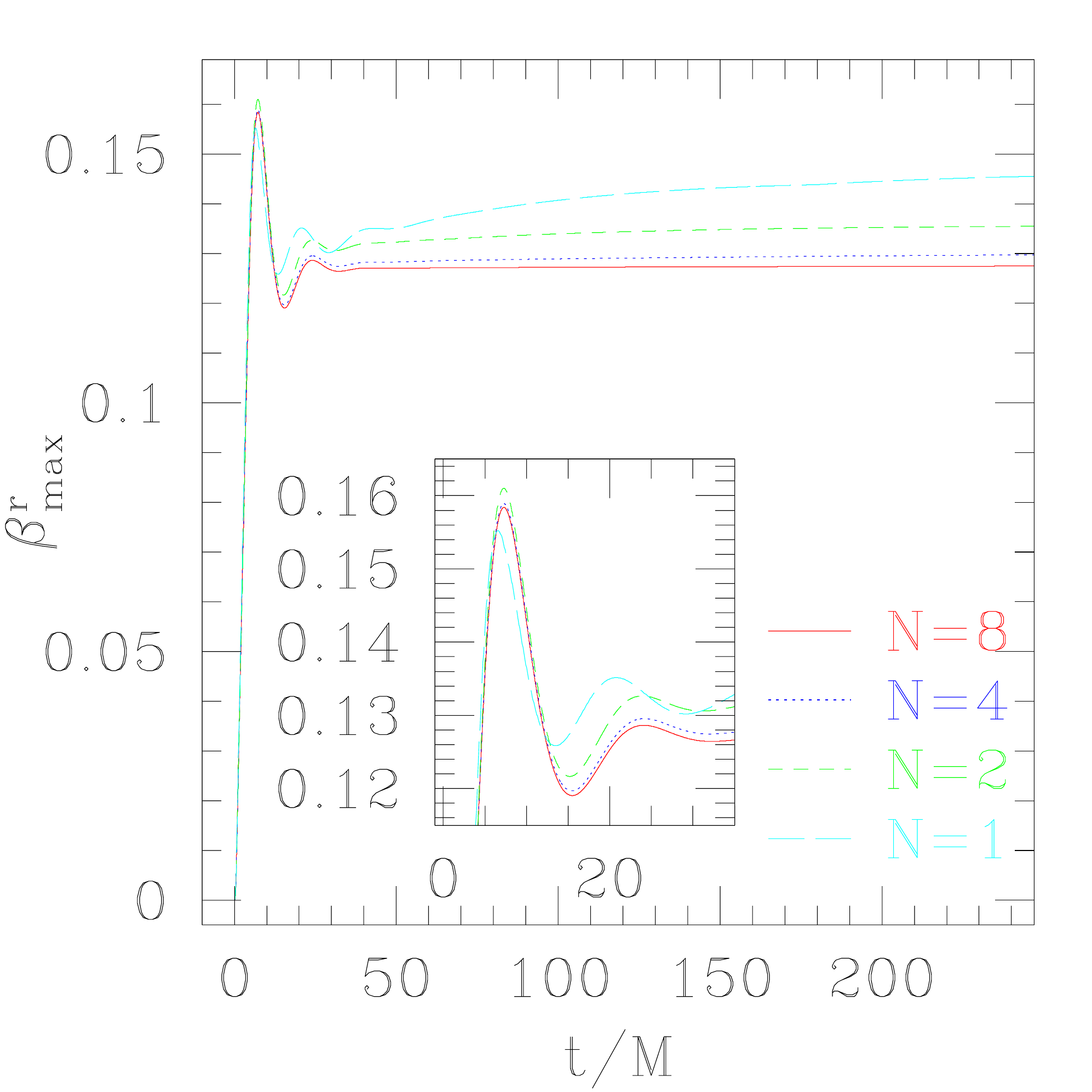}
\caption{The maximum of the radial shift $\beta^r$ as a function of time.  We show results for different grid sizes $(1280N,2,2)$ for $N=1$, 2, 4 and 8, with the outer boundary imposed at $256M$.  After a brief transition from the initial data $\beta^r = 0$, the shift settles down into a new equilibrium.  For relatively coarse grid resolutions the shift experiences a slow drift, but this drift disappears as the grid resolution is increased.} 
\label{Fig8}
\end{figure}

\begin{figure}[t]
\includegraphics[width=3.4in]{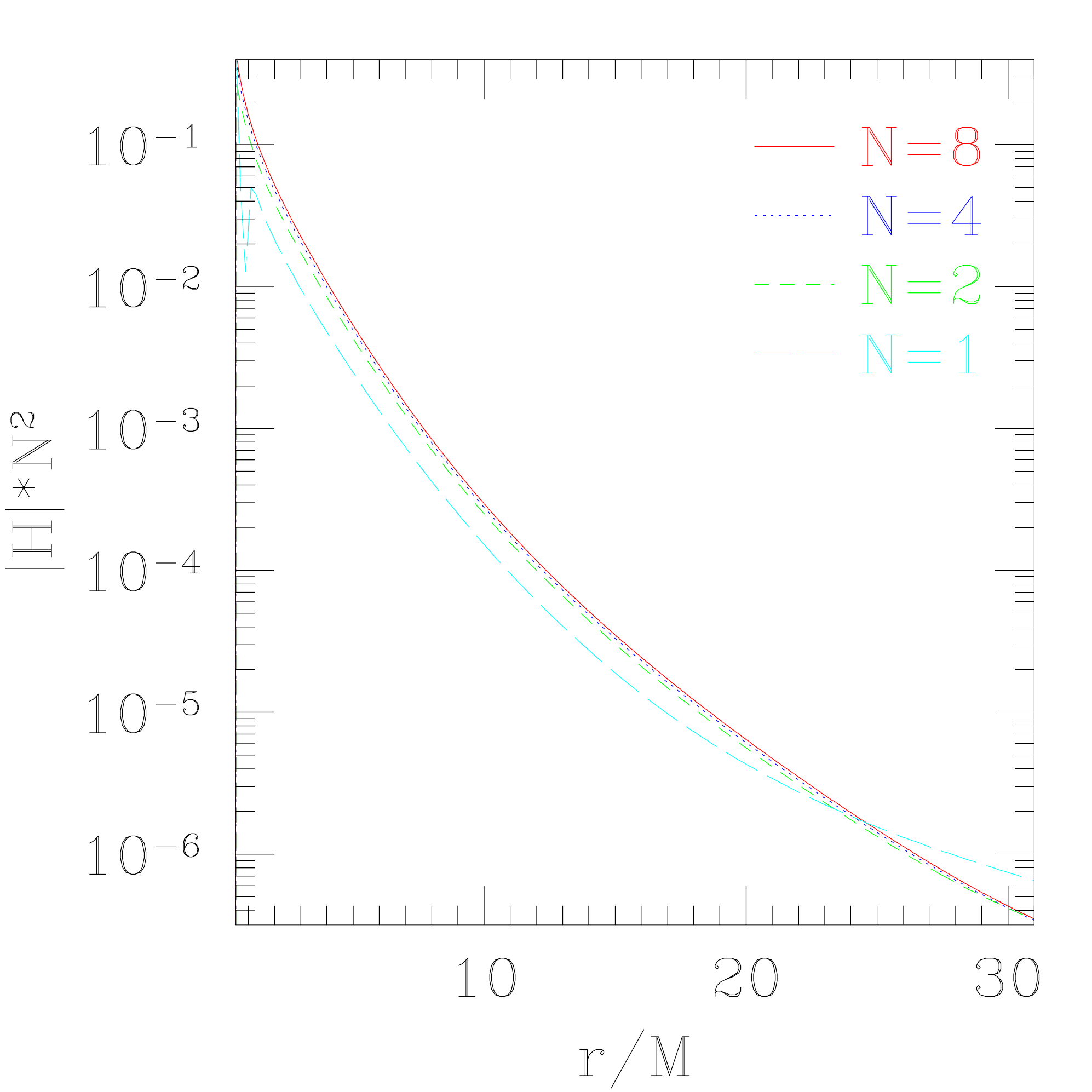}
\caption{Profiles of violations of the Hamiltonian constraint (\ref{Ham}) at time $t=79M$.  As in Fig.~\ref{Fig8} we show results for grid sizes $(1280N,2,2)$ for $N=1$, 2, 4 and 8, with the outer boundary imposed at $256M$.   All results are rescaled with $N^2$; the convergence of the resulting lines demonstrates second-order convergence of our code.}
\label{Fig9}
\end{figure}

We now turn to evolutions of wormhole initial data, representing a horizontal slice through the Penrose diagram of a Schwarzschild black hole.  For these data, the conformal factor is given by
\begin{equation}
\psi = 1 + \frac{M}{2r},
\end{equation}
the conformally related metric is flat, $\bar \gamma_{ij} = \eta_{ij}$, and the extrinsic curvature vanishes, $\bar A_{ij} = 0 = K$.   Instead of choosing the Killing lapse and Killing shift, which would leave these data time-independent, we choose, at $t=0$, a ``pre-collapsed" lapse \cite{AlcBDKPST03}
\begin{equation}
\alpha = \psi^{-2}
\end{equation}
and a vanishing shift, $\beta^i =0$.  We then evolve the lapse and the shift with the 1+log condition (\ref{1+log}) and the Gamma-driver condition (\ref{gammadriver}). 

Since these initial data do not represent a time-independent solution to the Einstein equations together with our gauge conditions, we observe a non-trivial time evolution that represents a coordinate evolution.  For the ``non-advective" 1+log condition (\ref{1+log}), this coordinate transition results in the maximally sliced trumpet geometry of Section \ref{sec:wormhole} \cite{HanHOBO08}.   In Fig.~\ref{Fig7} we show this coordinate transition for the conformal exponent $\phi$, the lapse $\alpha$ and the shift $\beta^r$.  

We note that some care has to be taken when the numerical and analytical results are compared.
The analytical solution of \cite{BauN07} assumes $\bar \gamma_{ij} = \eta_{ij}$.  We also choose $\bar \gamma_{ij} = \eta_{ij}$ in our initial data, but this relation is not necessarily maintained during the time evolution, so that the numerical and analytical solutions may be represented in different spatial coordinate systems (but on the same spatial slice).   In order to compare the two solutions we therefore graph all quantities as a function of the gauge-invariant areal radius $R$.  Since for wormhole data each value of $R>2M$ corresponds to two values of the isotropic radius $r$, the initial data in Fig.~\ref{Fig7} appear double-valued.  For these comparisons with the analytical solution we also graph the orthonormal component of the shift $\beta^{\hat r}$ rather than the coordinate component $\beta^r$ itself.  Fig.~\ref{Fig7} clearly shows the coordinate transition from wormhole initial data to the trumpet equilibrium solution.  

In Fig.~\ref{Fig8} we show the maximum of the radial shift $\beta^r$ as a function of time.  After a brief period of a coordinate transition the shift settles down into a new equilibrium.   We show results for grid sizes  $(1280N,2,2)$ for $N=1$, 2, 4 and 8, with the outer boundary imposed at $256M$.  The graph shows that differences between the different results decrease rapidly as the grid resolution is increased.   For our coarser grid resolutions the shift still experiences a slow drift after the initial transition, but this drift decreases as the grid resolution is increased.

Finally, in Fig.~\ref{Fig9}, we show profiles of the violations of the Hamiltonian constraint (\ref{Ham}) at time $t = 79M$.  In this graph all results are rescaled with $N^2$; the convergence of the resulting lines demonstrates that the numerical error in these simulations is again dominated by the second-order implementations of the advective shift terms, and possibly the time evolution.

\section{Discussion}
\label{sec:discussion}

In this paper we demonstrate that a PIRK method can be used to solve the Einstein equations in spherical polar coordinates without any need for any regularization at the origin or on the axis.  Specifically, we integrate a covariant version of the BSSN equations in three spatial dimensions without any symmetry assumptions.  To the best of our knowledge, these calculations represent the first successful three-dimensional numerical relativity simulations using spherical polar coordinates.  We consider several test cases to assess the stability, accuracy and convergence of the code, namely weak-field ``Teukolsky" gravitational waves, ``hydro-without-hydro" simulations of static and rotating relativistic stars, and single black holes.

Spherical polar coordinates have several advantages and disadvantages over Cartesian coordinates.  At least in single-grid calculations, spherical polar coordinates allow for a more effective allocation of the numerical grid points for applications that involve one center of mass, for example gravitational collapse of single stars or supernovae.  This is true even for uniform grids, which we adopt in this paper, but curvilinear coordinate systems also facilitate the use of non-uniform grids (e.g.~a logarithmic radial coordinate) to achieve a high resolution near the origin while keeping the outer boundary sufficiently far.  

Spherical polar coordinates have another strong advantage over Cartesian coordinates.  In simulations of supernovae or gravitational collapse, for example, the shape of the stellar objects is not well represented by Cartesian grids.  This mismatch between the symmetry of the object and the grid creates direction-dependent numerical errors, which are observed to trigger $m=4$ modes that grow in time.   Since spherical polar coordinates mimic the symmetry of collapsing stars more accurately, we expect that this problem can at least be reduced with these coordinates. 

However, spherical polar coordinates also have disadvantages.  One of these disadvantages is of practical nature: the equations in spherical polar coordinates include many more terms than those in Cartesian coordinates, which makes the numerical implementation more cumbersome and error prone.  Spherical polar coordinates also introduce coordinate singularities that traditionally have created many numerical problems; but these problems can be avoided when using a PIRK method.

Perhaps the most severe disadvantage of spherical polar coordinates is caused by the Courant limitation on the time step.  As shown in eq.~(\ref{Courant}), the close proximity of grid points close to the origin limits the size of the time steps $\Delta t$ to increasingly small values as the resolution is increased.  In three-dimensional simulations, $\Delta t$ decreases approximately with the product $\Delta t \propto \Delta r \Delta \theta \Delta \phi$.  This is a severe disadvantage compared to Cartesian coordinates where typically $\Delta t \propto \Delta x^i$.  However, this problem is not unique to numerical relativity, and instead is well-known from dynamical simulations in spherical polar coordinates in any field.  Accordingly, several different approaches to either solving or reducing this problem have been suggested.

One possible approach is to reduce the grid resolution in the angular directions, $N_\theta$ and $N_\phi$, close to the origin.  However, for many applications the angular dependence of the solution may be independent of the radius, so that this approach might severely limit the accuracy of the results.    It may also be possible to replace the PIRK method in a sphere around the origin with a completely implicit scheme, so that the time step there is no longer limited by the Courant condition (\ref{Courant}).  Similar implicit/explicit (IMEX) ``split-by-region'' schemes have been suggested, for example, in \cite{KanCGH07} in the context of spectral schemes.   Finally, the ``Yin-Yang" method suggested in \cite{WonHM10} mitigates the restrictions imposed by the Courant condition (\ref{Courant}) as follows.  Note that the smallest physical distance between grid points, which in turn limits the time step $\Delta t$, occurs next to the axis.  In the Yin-Yang method, the unit sphere is therefore covered by two different grids that are rotated by an angle of 90 degrees with respect to each other.  Each one covers only a region around its equator, thereby avoiding the most severe limitation on the time step next to the axis, but combined both grids cover the entire unit sphere.

Despite the small time step, however, we have been able to complete all simulations presented in this paper even with a serial code -- in fact, some of our simulations were performed on a laptop computer.

\acknowledgments

TWB and ICC gratefully acknowledge support from the Alexander-von-Humboldt Foundation, TWB would also like to thank the Max-Planck-Institut f\"ur Astrophysik for its hospitality.  This work was supported in part by the Deutsche Forschungsgemeinschaft (DFG) through its Transregional Center SFB/TR7 ``Gravitational Wave Astronomy'', and by NSF grant PHY-1063240 to Bowdoin College.

\begin{appendix}

\section{A numerical test for curvature quantities in spherical polar coordinates}
\label{appendixA}

In spherical polar coordinates, in particular in the absence of any symmetry assumptions, the numerical implementation of curvature quantities involves a significant number of terms that can easily introduce mistakes (see Section \ref{sec:implementation}).  One way of testing this part of the numerical code is to compare with known analytical solutions, for example for the Schwarzschild metric.  However, most analytical solutions feature symmetries (e.g.~spherical symmetry for Schwarzschild) that simplify the problem in the spherical polar coordinates of our code.  As a consequence, many terms vanish identically for these solutions, so that not all terms in the code are tested.  In this Appendix we describe a simple test that is also analytical, but is neither spherically nor axially symmetric, and hence a very stringent test.

Starting with the flat metric in Cartesian coordinates we introduce a coordinate transformation of each coordinate $x^i$ that only depends on that coordinate itself; the resulting metric then takes the form 
\begin{equation} \label{flat_metric}
\bar \gamma_{ij} = \left(
\begin{array}{ccc}
f(x)  & 0 & 0 \\
0 & g(y)  & 0 \\
0  & 0 & h(z) 
\end{array}
\right),
\end{equation}
where $f(x)$, $g(y)$ and $h(z)$ are arbitrary functions.  Transforming this metric into spherical polar coordinates leads to a metric for which, in general, all coefficients are non-zero and depend on the coordinates in potentially complicated ways.  

In Cartesian coordinates, the only non-vanishing Christoffel symbols are
\begin{eqnarray}
\Delta \Gamma^x_{xx} & = & \bar \Gamma^x_{xx} = \frac{f'(x)}{2f},\\
\Delta \Gamma^y_{yy} & = & \bar \Gamma^y_{yy} = \frac{g'(y)}{2g},\\
\Delta \Gamma^z_{zz} & = & \bar \Gamma^z_{zz} = \frac{h'(z)}{2h}
\end{eqnarray}
where the prime denotes a derivative with respect to the argument.  Given that the $\Delta \Gamma^i_{jk}$ transform like tensors, we can obtain the corresponding coefficients in spherical polar coordinates with a simple coordinate transformation.  For sufficiently general functions $f(x)$, $g(y)$ and $h(z)$, all 18 components of $\Delta \Gamma^i_{ij}$ in spherical polar coordinates will be non-zero.  This yields analytical expressions for the connection coefficients (\ref{connection}) that can then be compared with numerical results.

\begin{figure}[t]
\includegraphics[width=3.4in]{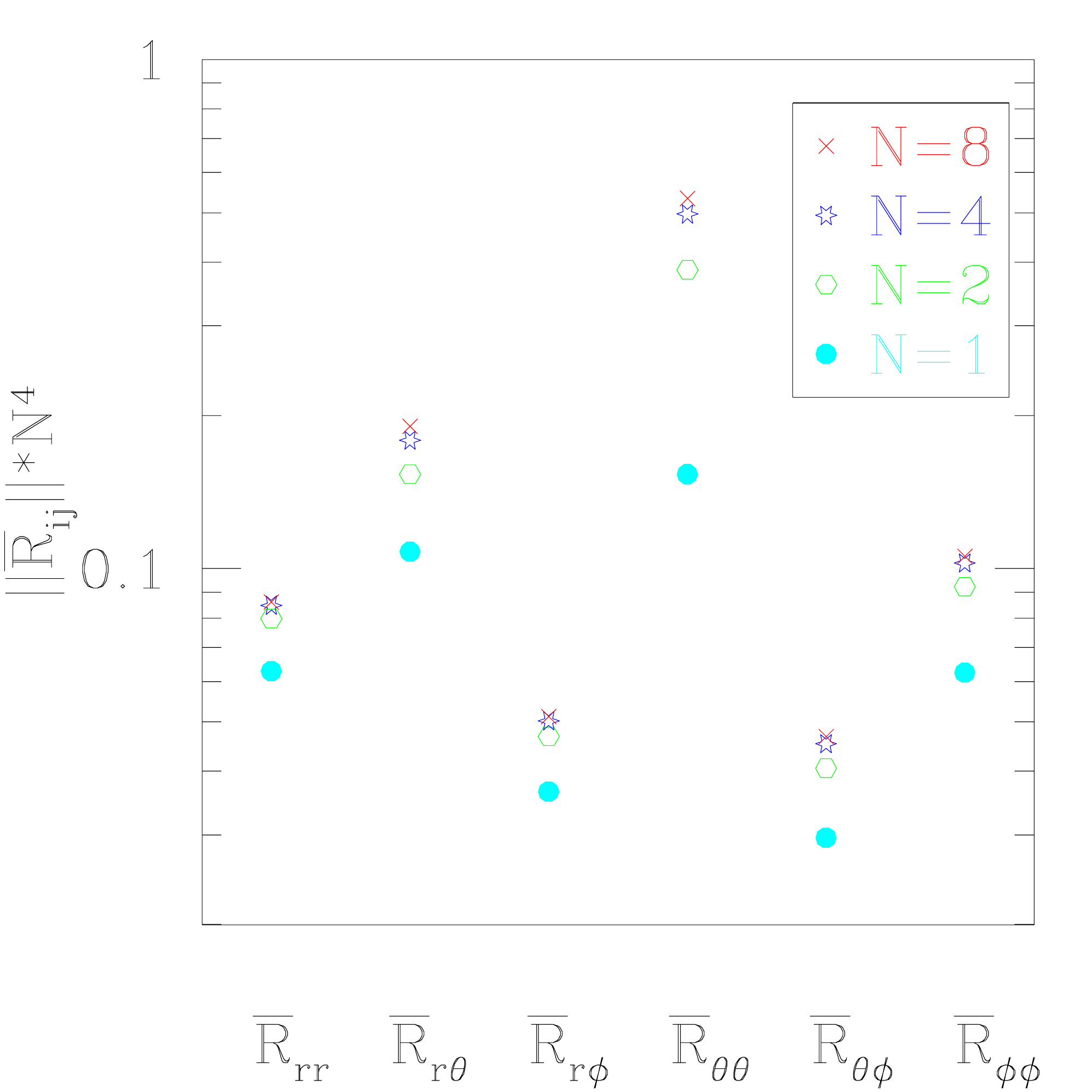}
\caption{Values of the norms of the components of the Ricci tensor, $||R_{ij}||$, for the flat metric (\ref{flat_metric}) with functions (\ref{flat_functions}), evaluated using grid sizes $(16N,8N,16N)$ for $N=1$, 2, 4 and 8.  All values are rescaled with $N^4$, so that the convergence of these results indicates fourth-order convergence of our implementation.}
\label{Fig10}
\end{figure}

Similarly, the connection functions are given by
\begin{equation}3.4in
\Lambda^i = \left( \frac{f'(x)}{2f^2}, \frac{g'(y)}{2 g^2}, \frac{h'(z)}{2h^2}  \right)
\end{equation}
in Cartesian coordinates, and can be transformed into spherical polar coordinates with a simple coordinate transformation.  

Finally, all components of the Ricci tensor in spherical polar coordinates should converge to zero, since the metric (\ref{flat_metric}) is still flat.  In Fig.~\ref{Fig10} we show numerical examples for 
\begin{eqnarray} \label{flat_functions}
f(x) & = & 1 + 0.1 \, x^2, \nonumber \\
g(y) & = & 1 + 0.3 \, y^2,\\
h(z) & = & 1 + 0.5 \, z^2. \nonumber
\end{eqnarray}
All components of $\bar R_{ij}$ are non-zero, but converge to zero as the grid resolution is increased.   In the graph we rescale all results with $N^4$, so that the convergence of the resulting quantities indicates fourth-order convergence of our implementation of the Ricci tensor, as expected.

\section{Detailed source terms included in the PIRK operators for the evolution
equations}
\label{appendixB}

We evolve the evolution eqs.~(\ref{evolution}),
(\ref{1+log})-(\ref{gammadriver}) using a second-order PIRK method.
In this Appendix we provide details on how we split the right-hand
sides of these equations into the explicit and partially implicit
operators.

We start each time step by evolving the conformal metric components,
$h_{ij}$, the conformal factor $\phi$, the lapse function, $\alpha$,
and the shift vector, $\beta^i$, explicitly, i.e., all the
source terms of the evolution equations of these variables are
included in the $L_1$ operator of the second-order PIRK method.

We then evolve the traceless part of the extrinsic curvature, $a_{ij}$, and
the trace of the extrinsic curvature, $K$, partially
implicitly. More specifically, the corresponding $L_2$ and $L_3$
operators associated with the evolution equations for $a_{ij}$ and $K$
in terms of the original BSSN variable $\bar A_{ij}$, related to
$a_{ij}$ through eq.~(\ref{Acap}), are
\begin{align}
        L_{2(\bar A_{ij})} &= e^{- 4 \phi} \Big[ - 2 \alpha \bar D_i \bar D_j \phi + 4 \alpha \bar D_i \phi \bar D_j \phi \nonumber \\
&  + 4 \bar D_{(i} \alpha \bar D_{j)} \phi - \bar D_i \bar D_j \alpha + \alpha \bar R_{ij} \Big]^{\rm TF}, \label{L2A} \\
        L_{3(\bar A_{ij})} &= - \frac{2}{3} \bar A_{ij} \bar D_k \beta^k - 2 \alpha \bar A_{ik} \bar A^k {}_j 
+ \alpha \bar A_{ij} K, \\
        L_{2(K)} &= - e^{- 4 \phi}  ( \bar D^2 \alpha + 2 \bar D^i
        \alpha \bar D_i \phi ) + \alpha \bar A_{ij} \bar A^{ij}, \\
        L_{3(K)} &=  \frac{\alpha}{3} K^2.
\end{align} 
The $\lambda^{i}$ are evolved partially implicitly, 
using the updated values of $\alpha$,
$\beta^i$, $\phi$, $h_{ij}$, $a_{ij}$ and  $K$. In terms of the
original BSSN variable $\bar \Lambda^i$, related to
$\lambda^{i}$ through eq.~(\ref{lambda}), the operators are
\begin{align}
        L_{2(\bar \Lambda^i)} &= \bar \gamma^{jk} \Dflat_j \Dflat_k \beta^i 
+  \frac{1}{3} \bar D^i \bar D_j \beta^j - \frac{4}{3} \alpha \bar
\gamma^{ij} \partial_j K \nonumber \\
&  - 2 \bar A^{jk} ( \delta^i{}_j \partial_k \alpha - 6 \alpha \delta^i{}_j \partial_k \phi
- \alpha \Delta \Gamma^i_{jk} ),\\   
	\label{L3Lambda}
        L_{3(\bar \Lambda^i)} &= \frac{2}{3} \Delta \Gamma^i \bar D_j \beta^j .
\end{align}
We note that the evaluation of the Ricci tensor $\bar R_{ij}$ in eq.~(\ref{L2A}) requires updated values of $\Lambda^i$ before they become available.  It is possible to either replace these updated values with old values, or to update the $\Lambda^i$ provisionally in a purely explicit step, use these values in eq.~(\ref{L2A}), but then overwrite these values after the $\Lambda^i$ are updated partially implicitly.  We have used the latter approach in the simulations presented in this paper.

Finally, the $B^i$ are evolved partially implicitly, using the 
updated values of the previous quantities,
\begin{align}
        L_{2(B^i)} &= \frac{3}{4}\partial_{t}\bar \Lambda^i, \\
        L_{3(B^i)} &= 0.
\end{align}
Matter source terms and Lie derivative terms are always included in
the explicitly treated parts.

\end{appendix}

\end{document}